\documentclass[twocolumn,superscriptaddress,showpacs,preprintnumbers,amsmath,amssymb]{revtex4}

\usepackage{graphicx}
\usepackage{dcolumn}
\usepackage{bm}

\begin{document}

\title{Probing non-tensorial polarizations of stochastic gravitational-wave backgrounds \\with ground-based laser interferometers}

\author{Atsushi~Nishizawa}
\email{atsushi.nishizawa@nao.ac.jp}
\affiliation{Graduate School of Human and Environmental Studies, Kyoto University, Kyoto 606-8501, Japan}
\author{Atsushi~Taruya}
\affiliation{Research Center for the Early Universe, School of Science, The University of Tokyo, Tokyo 113-0033, Japan}
\affiliation{Institute for the Physics and Mathematics of the Universe, University of Tokyo, Kashiwa, Chiba 277-8568, Japan}
\author{Kazuhiro~Hayama}
\affiliation{
Albert-Einstein-Institut (Max-Planck-Institut f\"ur Gravitationsphysik), Callinstra\ss e 38, D-30167 Hannover, Germany}
\author{Seiji~Kawamura}
\affiliation{TAMA Project, National Astronomical Observatory of Japan, Mitaka, Tokyo 181-8588, Japan}
\author{Masa-aki~Sakagami}
\affiliation{Graduate School of Human and Environmental Studies, Kyoto University, Kyoto 606-8501, Japan}

\date{\today}

\begin{abstract}
In a general metric theory of gravitation in four dimensions, six polarizations of a gravitational wave are allowed: two scalar and two vector modes, in addition to two tensor modes in general relativity. Such additional polarization modes appear due to additional degrees of freedom in modified theories of gravitation or theories with extra dimensions. Thus, observations of gravitational waves can be utilized to constrain the extended models of gravitation. In this paper, we investigate detectability of additional polarization modes of gravitational waves, particularly focusing on a stochastic gravitational-wave background, with laser-interferometric detectors on the Earth. We found that more than three detectors can separate the mixture of polarization modes in detector outputs, and that they have almost the same sensitivity to each polarization mode of stochastic gravitational-wave background.
\end{abstract}
\maketitle

\section{Introduction}
In recent years, direct detection experiments of a gravitational wave (GW) have been well developed. The first generation of a kilometer-scale ground-based laser-interferometric GW detector, located in the United States (LIGO), Europe (VIRGO and GEO 600), and Japan (TAMA 300), has begun its search for gravitational-waves and has yielded scientific results: especially for a stochastic gravitational-wave background (GWB), the upper limit on the energy density has been obtained \cite{bib38,bib39,bib40}. However, the first detection of the GW has not been achieved yet. The development of interferometers of the next generation, such as AIGO \cite{bib13}, advanced LIGO \cite{bib14}, advanced VIRGO \cite{bib15}, and LCGT \cite{bib12}, is underway. In the coming decade, a direct detection of GWs will be made, and the GW experiments will be a key observational tool to obtain valuable information about astronomical objects and physics of the early universe. 

The direct observation of the GWs will also provide a unique opportunity to test the theory of general relativity (GR), through the propagation speed, waveforms, and polarization modes of GWs. In GR, a GW has two polarization modes (plus and cross modes), while in a general metric theory of gravitation, the GW is allowed to have, at most, six polarizations \cite{bib1,bib2}. In the extended theory of gravitation such as Brans-Dicke theory \cite{bib29,bib30} and $f(R)$ gravity \cite{bib35}, the GW has additional polarization modes, because of extra degrees of freedom involved with scalar fields. In the theories with extra dimensions such as the Kaluza-Klein theory and the Dvali-Gabadadze-Porrati (DGP) braneworld model \cite{bib34}, gravitons are able to propagate into extra dimensions, and have additional polarizations. If additional polarizations are found, it indicates that the theory of gravitation should be extended beyond GR, and excludes some theoretical models, depending on which polarization modes are detected. Thus, the observation of the GW polarizations is a powerful tool to probe the extended law of gravity and extra dimensions. Subsequently, we can also investigate the accelerated expansion of the universe \cite{bib27,bib28}. 

Currently, there are few observational constraints on the additional polarization modes of GWs. For the scalar GWs, the observed orbital-period derivative of PSR B1913+16 agrees well with predicted values of GR, conservatively, at a level of $1\,\%$ error \cite{bib41}, indicating that the contribution of scalar GWs to the energy loss is less than $1\,\%$.
Another constraint on a scalar GWB has been obtained from the amplitude of scalar perturbations in the WMAP data, which corresponds to $h_0^2 \Omega_{\rm{gw}}^{S} \lesssim 10^{-12}$ \cite{bib42,bib26}, where the critical density of the universe is $\rho_c = 3 H_0^2/8\pi G$ and the Hubble constant is $H_0=100\,h_{0}\,\rm{km\, s^{-1}\,Mpc^{-1} }$. 

On the other hand, little work on a direct search has been done so far, except for the recent work by Lee, Jenet, and Price \cite{bib4}. The authors have investigated the detectability of non-Einsteinian (non-GR) polarizations of a GWB at the frequencies, $\sim 10^{-8}\,\rm{Hz}$, with pulsar timing, and showed that the non-Einsteinian polarizations can be discriminated with 40 - 60 stable pulsars. In this paper, using multiple laser-interferometric GW detectors on the Earth (at $\sim 100\,\rm{Hz}$), we present a method for separating a mixture of the polarization modes of the GWB and detecting non-tensorial polarization modes.  

This paper is organized as follows. In Sec.\,\ref{sec2}, we define the six polarization modes of a GW. In Sec.\,\ref{sec3}, we investigate the response of a single detector to the GW propagating in a certain direction with each polarization mode. In Sec.\,\ref{sec4}, we focus on a GWB, and describe cross-correlation between two detectors, extending the analysis to the non-tensorial polarizations. Then, we discuss an optimal location and orientation of the detectors, and the detectability of the GWB with a single polarization mode. In Sec.\,\ref{sec5}, we consider the separation of a polarization mode using multiple detectors, which is the main subject of this paper, and estimate the detector sensitivity to the GWB for various combinations of multiple detectors. Finally, Sec.\,\ref{sec6} is devoted to conclusions and discussions for future prospects.

\begin{figure*}[t]
\begin{center}
\includegraphics[width=14cm]{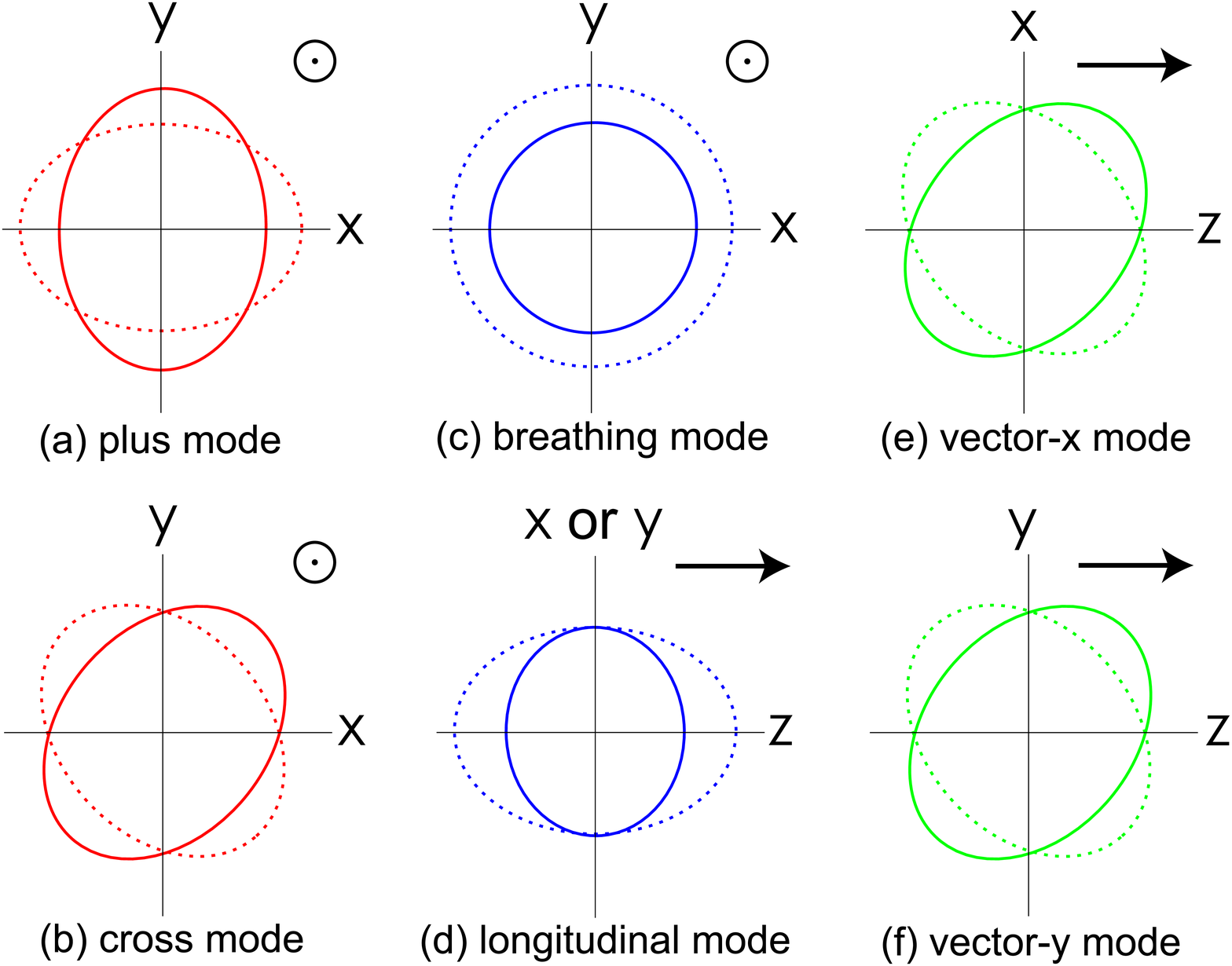}
\caption{(color online). Six GW polarizations in a general metric theory of gravitation. The two ellipses (or circles) show the effect of a GW with each polarization on test masses arranged on a circle at the moments of different phases by $\pi$. The symbol $\bigodot$ and the arrow represent the propagating direction of the GW.}
\label{fig1}
\end{center}
\end{figure*}

\section{GW polarization modes}
\label{sec2}
In general, a metric gravity theory in four dimensions allows, at most, six polarization modes of a GW \cite{bib1,bib2}. If a spacetime includes extra dimensions, the number of polarization modes can be more than six. However, once the polarizations are projected onto our 3-space, the polarizations we observe are degenerate, and are classified into six polarizations. For the GW propagating in the $z$ direction, the bases of the six polarizations are defined by ($x,y$, and $z$ components are from the left to the right or the top to the bottom in the tensors.)
\begin{eqnarray}
\tilde{e}_{ij}^{+}&=& 
\left(
\begin{array}{ccc} 
1 & 0 & 0  \\
0 & -1 & 0  \\
0 & 0 & 0 
\end{array}
\right)\;, \;\;\;
\tilde{e}_{ij}^{\times}= 
\left(
\begin{array}{ccc} 
0 & 1 & 0  \\
1 & 0 & 0  \\
0 & 0 & 0 
\end{array}
\right)\;, \nonumber \\
\tilde{e}_{ij}^{b}&=& 
\left(
\begin{array}{ccc} 
1 & 0 & 0  \\
0 & 1 & 0  \\
0 & 0 & 0 
\end{array}
\right)\;, \;\;\;
\tilde{e}_{ij}^{\ell}= \sqrt{2}
\left(
\begin{array}{ccc} 
0 & 0 & 0  \\
0 & 0 & 0  \\
0 & 0 & 1 
\end{array}
\right)\;, \nonumber \\ 
\tilde{e}_{ij}^{x} &=&
\left(
\begin{array}{ccc} 
0 & 0 & 1  \\
0 & 0 & 0  \\
1 & 0 & 0 
\end{array}
\right)\;, \;\;\;
\tilde{e}_{ij}^{y}= 
\left(
\begin{array}{ccc} 
0 & 0 & 0  \\
0 & 0 & 1  \\
0 & 1 & 0 
\end{array}
\right)\;, 
\label{eq1} 
\end{eqnarray}
where $+$, $\times$, $b$, $\ell$, $x$, and $y$ denote plus, cross, breathing, longitudinal, vector-x, and vector-y modes, respectively. The tildes are fixed to represent that the polarizations are defined in our 3-space, not in a spacetime with extra dimensions. Each polarization mode is orthogonal to one another and is normalized so that $\tilde{e}_{ij}^{A} \tilde{e}^{ij}_{A^{\prime}}=2 \delta _{AA^{\prime}}, \; A,A^{\prime}=+,\times,b,\ell,x,$ and $y$. Note that the breathing and longitudinal modes are not traceless, in contrast to the ordinary plus and cross polarization modes in GR. In Fig.\,\ref{fig1}, we illustrate how each GW polarization affects test masses arranged on a circle. According to rotation symmetry around the propagation axis of the GW, the $+$ and $\times$ modes can be identified with tensor-type (spin-2) GWs, the $x$ and $y$ modes are vector-type (spin-1) GWs, and the $b$ and $\ell$ modes are scalar-type (spin-0) GWs. 

The polarization modes in various alternative theories are summarized in Table \ref{tab1}. All polarizations do not necessarily appear in all theoretical models. Most of the gravity theories in cosmologically interesting situations are the variant of the scalar-tensor theory, where no vector polarization appears. 

\begin{table*}[t]
\begin{center}
\begin{tabular}{l|c|c|c|c|c|c|c}
\hline
\hline
theoretical model \quad \quad \quad \quad \quad \quad \quad \quad \quad \quad \quad\quad \quad \quad & $\;\; \tilde{e}_{ij}^{+} \;\;$ & $\;\; \tilde{e}_{ij}^{\times} \;\;$ & $\;\; \tilde{e}_{ij}^{b} \;\;$ & $\;\; \tilde{e}_{ij}^{\ell} \;\;$ & $\;\; \tilde{e}_{ij}^{x} \;\;$ & $\;\; \tilde{e}_{ij}^{y} \;\;$ & \;\; references \;\; \\
\hline
GR in a noncompactified 5D Minkowski spacetime & $\bigcirc$ & $\bigcirc$ & $\bigcirc ^1$ & $\bigcirc ^1$ & $\bigcirc$ & $\bigcirc$ & --- \\ 
GR in a noncompactified 6D Minkowski spacetime & $\bigcirc$ & $\bigcirc$ & $\bigcirc$ & $\bigcirc$ & $\bigcirc$ & $\bigcirc$ & --- \\ 
5D Kaluza-Klein theory & $\bigcirc$ & $\bigcirc$ & $\bigcirc$ & --- & $\bigcirc$ & $\bigcirc$ & \cite{bib18} \\
Randall-Sundrum braneworld & $\bigcirc$ & $\bigcirc$ & --- & --- & --- & --- & \cite{bib19}  \\
DGP braneworld (normal branch) & $\bigcirc$ & $\bigcirc$ & --- & --- & --- & --- & \cite{bib20}  \\
DGP braneworld (self-accelerating branch) & $\bigcirc$ & $\bigcirc$ &  $\bigcirc^2$ & $\bigcirc^2$ & $\bigcirc$ & $\bigcirc$ & \cite{bib20} \\
\hline
Brans-Dicke theory & $\bigcirc$ & $\bigcirc$ & $\bigcirc^2$ & $\bigcirc^2$ & --- & --- & \cite{bib23,bib22}  \\
$f(R)$ gravity & $\bigcirc$ & $\bigcirc$ & $\bigcirc^2$ & $\bigcirc^2$ & --- & --- & \cite{bib25,bib26}  \\
Bimetric theory & $\bigcirc$ & $\bigcirc$ & $\bigcirc^2$ & $\bigcirc^2$ & $\bigcirc$ & $\bigcirc$  & \cite{bib24} \\
\hline 
\hline
\end{tabular}
\end{center}
\caption{GW polarization modes in various theories. $^1$In a general five-dimensional spacetime, two scalar modes are correlated and behave as one degree of freedom. $^2$In the case that graviton is massless ($m_g=0$), the longitudinal mode vanishes. On the other hand, in the case of massive graviton ($m_g\neq 0$), the breathing and longitudinal modes are correlated.}
\label{tab1}
\end{table*}

\section{Response of a single GW detector}
\label{sec3}
In this section, as preliminaries of later analysis, we consider the response of a single detector to a GW propagating in a certain direction. A perturbed metric $h_{ij}$, which represents the GW propagating in three-dimensional space, is decomposed into the six polarization modes as 
\begin{eqnarray}
h_{ij}(\omega t-\vec{k} \cdot \vec{x})& =& h_{+}(\omega t-\vec{k} \cdot \vec{x}) \tilde{e}_{ij}^{+} + h_{\times}(\omega t-\vec{k} \cdot \vec{x}) \tilde{e}_{ij}^{\times} \nonumber \\
&+& h_b(\omega t-\vec{k} \cdot \vec{x}) \tilde{e}_{ij}^{b} +h_{\ell} (\omega t-\vec{k} \cdot \vec{x}) \tilde{e}_{ij}^{\ell} \nonumber \\
&+& h_x (\omega t-\vec{k} \cdot \vec{x}) \tilde{e}_{ij}^{x}+h_y(\omega t-\vec{k} \cdot \vec{x}) \tilde{e}_{ij}^{y} \;, \nonumber
\end{eqnarray}
where $h_A,\, A=+,\times,b,\ell,x,$ and $y$ are the amplitudes of GWs for each mode.

\begin{figure}[h]
\begin{center}
\includegraphics[width=6.5cm]{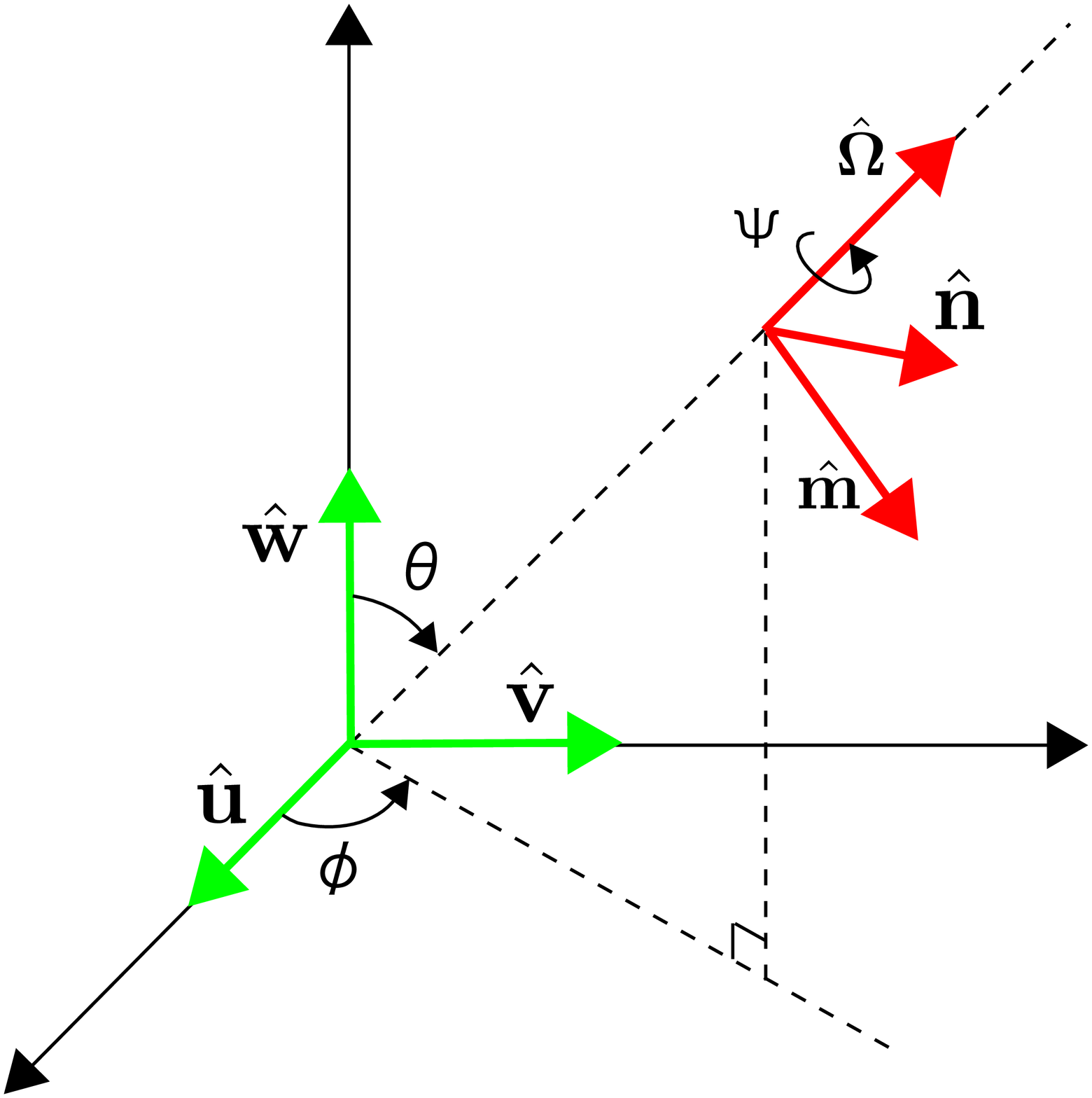}
\caption{(color online). Coordinate systems.}
\label{fig2}
\end{center}
\end{figure}

Although a detector output actually depends on the GW amplitude determined by a specific theoretical model, we can discuss the detector response to each GW polarization without specifying a certain theoretical model. The angular pattern function of a detector to GWs is given by
\begin{eqnarray}
F_A (\hat{\mathbf{\Omega}}) &=& \mathbf{D} : \tilde{\mathbf{e}}_A
(\hat{\mathbf{\Omega}})\:, 
\label{eq2} \\
\mathbf{D} &=&  \frac{1}{2}\left[ \hat{\mathbf{u}} \otimes \hat
{\mathbf{u}}- \hat{\mathbf{v}}
\otimes \hat{\mathbf{v}}\right]\:,
\nonumber
\end{eqnarray}
where the symbol : denotes contraction between tensors, and $\mathbf{D}$ is a so-called detector tensor, which 
describes the response of a laser-interferometric detector and maps the gravitational metric perturbation
to a GW signal from the detector. The unit vectors $\hat{\mathbf{u}}$ and $\hat{\mathbf{v}}$ are orthogonal to each other and are directed to each detector arm, which form an orthonormal coordinate system with the unit vector $\hat{\mathbf{w}}$, as shown in Fig.\,\ref{fig2}. $\hat{\mathbf{\Omega}}$ is the unit vector directed at the GW propagation direction. Note that the detector tensor, Eq.\,(\ref{eq2}), is valid only when the arm length of the detector is much smaller than the wavelength of GWs that we consider. This is relevant for our purpose to deal with the ground-based laser interferometers.

Suppose that an orthonormal coordinate system for the detector is 
\begin{equation}
\left\{
\begin{array}{lll} 
\displaystyle
\hat{\mathbf{u}} = (1,0,0)
\\ 
\displaystyle 
\hat{\mathbf{v}} = (0,1,0)
\\ 
\displaystyle 
\hat{\mathbf{w}} = (0,0,1)
\end{array}
\right. \;, \nonumber
\end{equation}
and the GW coordinate system rotated by angles $(\theta, \phi)$ is
\begin{equation}
\left\{
\begin{array}{lll} 
\displaystyle
\hat{\mathbf{u}}^{\prime} =(\cos \theta \cos \phi , \cos \theta \sin \phi , -\sin \theta)
\\ 
\displaystyle 
\hat{\mathbf{v}}^{\prime} = (- \sin \phi , \cos \phi , 0)
\\ 
\displaystyle 
\hat{\mathbf{w}}^{\prime} = (\sin \theta \cos \phi , \sin \theta \sin \phi , \cos \theta)
\end{array}
\right. \;. \nonumber
\end{equation}
The most general choice of the coordinates is obtained by the rotation with respect to the angle $\psi$ around the GW-propagating axis,
\begin{equation}
\left\{
\begin{array}{lll} 
\displaystyle
\hat{\mathbf{m}} = \hat{\mathbf{u}}^{ \prime} \cos \psi + \hat{\mathbf{v}}^{\prime} \sin \psi\\ 
\displaystyle 
\hat{\mathbf{n}} = - \hat{\mathbf{v}}^{ \prime} \sin \psi + \hat{\mathbf{u}} ^{\prime} \cos \psi
\\ 
\displaystyle 
\hat{\mathbf{\Omega}} = \hat{\mathbf{w}}^{ \prime}
\end{array}
\right. \;. 
\nonumber
\end{equation}
The coordinate system $(\hat{\mathbf{u}},\hat{\mathbf{v}},\hat{\mathbf{w}})$ is related to the coordinate system $(\hat{\mathbf{m}},\hat{\mathbf{n}},\hat{\mathbf{\Omega}})$ by the rotation angles ($\phi,\,\theta,\,\psi$), shown in Fig.\,\ref{fig2}. Using the unit vectors $\hat{\mathbf{m}}$, $\hat{\mathbf{n}}$, and $\hat{\mathbf{\Omega}}$, the polarization tensors in Eq.\,(\ref{eq1}) can be written as
\begin{eqnarray}
\tilde{\mathbf{e}}_{+} &=& \hat{\mathbf{m}} \otimes \hat{\mathbf{m}} -\hat{\mathbf{n}} \otimes \hat{\mathbf{n}} \;, \nonumber \\
\tilde{\mathbf{e}}_{\times} &=& \hat{\mathbf{m}} \otimes \hat{\mathbf{n}} +\hat{\mathbf{n}} \otimes \hat{\mathbf{m}} \;, \nonumber \\
\tilde{\mathbf{e}}_{b} &=& \hat{\mathbf{m}} \otimes \hat{\mathbf{m}} + \hat{\mathbf{n}} \otimes \hat{\mathbf{n}} \;,  \nonumber \\
\tilde{\mathbf{e}}_{\ell} &=& \sqrt{2}\, \hat{\mathbf{\Omega}} \otimes \hat{\mathbf{\Omega}} \;,  \nonumber \\
\tilde{\mathbf{e}}_{x} &=& \hat{\mathbf{m}} \otimes \hat{\mathbf{\Omega}} +\hat{\mathbf{\Omega}} \otimes \hat{\mathbf{m}} \;, \nonumber \\
\tilde{\mathbf{e}}_{y} &=& \hat{\mathbf{n}} \otimes \hat{\mathbf{\Omega}} +\hat{\mathbf{\Omega}} \otimes \hat{\mathbf{n}} \;. \nonumber
\end{eqnarray}

\begin{figure*}[t]
\begin{center}
\includegraphics[width=14cm]{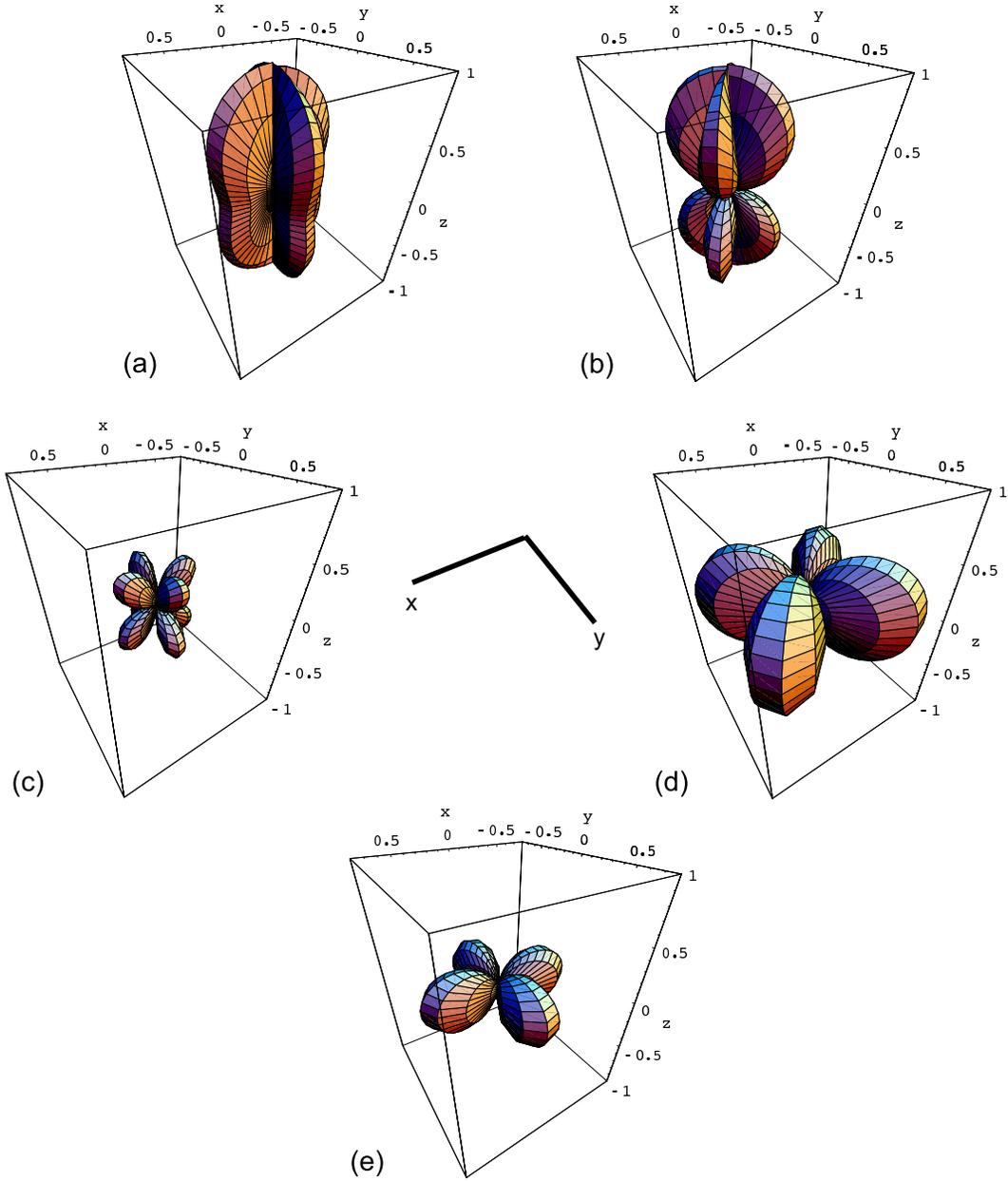}
\caption{(color online). Angular pattern functions of a detector for each polarization. (a) Plus mode $|F_+|$, (b) cross mode $|F_{\times}|$, (c) x mode $|F_{x}|$, (d) y mode $|F_{y}|$, and (e) longitudinal mode $|F_{\ell}|$. The angular pattern function of the breathing mode is the same as that of the longitudinal mode except for an overall factor $1/\sqrt{2}$. At the center of the figure, the arms of an interferometer are shown.}
\label{fig3}
\end{center}
\end{figure*}

\begin{figure*}[t]
\begin{center}
\includegraphics[width=16cm]{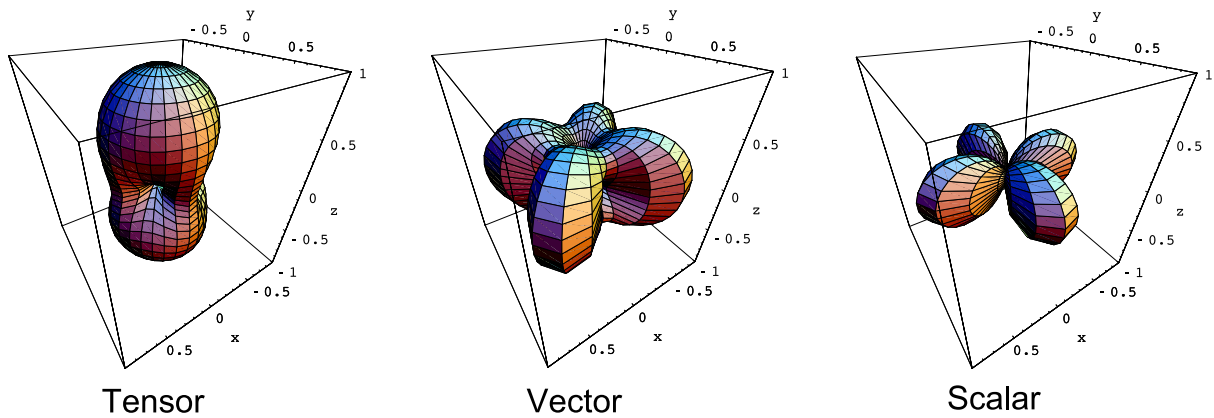}
\caption{(color online). Angular pattern functions of a detector for the tensor, vector, and scalar modes. The plots from the left are $\sqrt{F_{+}^2+F_{\times}^2}$,\, $\sqrt{F_{x}^2+F_{y}^2}$, and $\sqrt{F_{b}^2+F_{\ell}^2}$, respectively.}
\label{fig4}
\end{center}
\end{figure*}

Then, from Eqs.\,(\ref{eq2}), the angular pattern functions for each polarization result in 
\begin{eqnarray}
F_{+}(\theta, \phi, \psi) &=& \frac{1}{2} (1+ \cos ^2 \theta ) \cos 2\phi \cos 2 \psi \nonumber \\
&& - \cos \theta \sin 2\phi \sin 2 \psi \;, 
\label{eq4} \\
F_{\times}(\theta, \phi, \psi) &=& -\frac{1}{2} (1+ \cos ^2 \theta ) \cos 2\phi \sin 2 \psi \nonumber \\
&& -  \cos \theta \sin 2\phi \cos 2 \psi \;, 
\end{eqnarray}
\begin{eqnarray}
F_{x}(\theta, \phi, \psi) &=& \sin \theta \,(\cos \theta \cos 2 \phi \cos \psi -\sin 2\phi \sin \psi) \;, \nonumber \\
&&  \\
F_{y}(\theta, \phi, \psi) &=& - \sin \theta \,(\cos \theta \cos 2 \phi \sin \psi +\sin 2\phi \cos \psi) \;, \nonumber \\
&& \\
F_{b}(\theta, \phi) &=& -\frac{1}{2} \sin^2 \theta \cos 2\phi \;,  \\
F_{\ell}(\theta, \phi) &=& \frac{1}{\sqrt{2}} \sin^2 \theta \cos 2\phi \;. 
\label{eq5} 
\end{eqnarray}
From the dependence on the angle $\psi$, the $+$ and $\times$ modes are tensor type (spin-2), the $x$ and $y$ modes are vector type (spin-1), and the $b$ and $\ell$ modes are scalar type (spin-0). The angular pattern functions of the breathing and longitudinal modes are completely degenerated, which prohibits one to decompose the two scalar modes with a laser-interferometric GW detector. We plot the angular pattern functions for each non-tensorial polarization in Fig.\,\ref{fig3}, and the angular pattern functions for the tensor, vector, and scalar modes in Fig.\,\ref{fig4}. These results are consistent with those obtained in \cite{bib3,bib36,bib37}.

\section{Cross correlation}
\label{sec4}
We focus on a stochastic GWB \footnote{The non-tensorial polarizations of a GW from a point source on the sky are also an interesting subject that should be addressed, though we do not consider them in this paper.}, based on the detector responses obtained in the previous section, and discuss cross correlation between a pair of detectors. 


\subsection{Correlation analysis}
To distinguish the GWB signal from stochastic detector noise independent in each detector, one has to correlate signals between two detectors. The correlation analysis has been well developed by several authors \cite{bib5,bib6,bib7}. In this section, we extend the method to the non-tensorial polarizations.

At a position $\vec{\mathbf{X}}$, the gravitational metric perturbations in our 3-space are given by
\begin{eqnarray}
\mathbf{h}(t,\vec{\mathbf{X}}) &=& \sum_A \int _{S^2} d\hat{\mathbf{\Omega}} \int_{-\infty}^
{\infty}df\, \tilde{h}_A (f, \hat{\mathbf{\Omega}})\, \nonumber \\
&& \times e^{2\pi if (t-\hat{\mathbf{\Omega}} \cdot 
\vec{\mathbf{X}}/c)}\, \tilde{\mathbf{e}}_A(\hat{\mathbf{\Omega}})\:,  \label{eq6}
\end{eqnarray}
where $\tilde{h}_A (f, \hat{\mathbf{\Omega}})$ is the Fourier transform of the GW amplitude with polarizations $A=+, \times, b, \ell, x,$ and $y$. In Eq.\,(\ref{eq6}), we assumed that gravitons propagate with the speed of light in our 3-space. Strictly speaking, gravitons propagate with the speed less than that of light, if they are massive in some modified gravity theories or are projected onto our 3-space in the presence of extra dimensions. However, the mass is constrained by several observations of the galaxy \cite{bib9}, the solar system \cite{bib10}, and binary pulsars \cite{bib11}. The limits from the galaxy and the solar system are obtained from the observation in static gravitational fields, while the limit of binary pulsars comes from the change of the orbital period of the binary pulsars (PSR B1913+16 and PSR B1534+12), in dynamical gravitational fields (The limit from the galaxy is the most stringent, but may be less robust.). Thus, here we adopt the binary pulsar bound. Finn and Sutton \cite{bib11} considered energy loss from the binary system by emission of massive gravitons, and obtained the limit on the mass of gravitons, $(m_g/\omega_{\rm orbit})^2 < 0.003$, where $\omega_{\rm orbit}$ is the orbital frequency of the binary. This limit implies 
\begin{equation}
\frac{v_g}{c} = \sqrt{1-\left( \frac{m_g}{\omega_{\rm orbit}} \right)^2} \gtrsim 0.998 \;, \nonumber 
\end{equation}
The gravitons cannot change their speed by more than 0.2\% from the speed of light. Thus, setting $v_g=c$ does not affect cross-correlation analysis qualitatively. Hereafter we set $v_g=c$.

Using Eqs.\,(\ref{eq6}) and (\ref{eq2}), GW signal $h(t)$ can be 
written as
\begin{eqnarray}
h(t,\vec{\mathbf{X}}) &=& \sum_A \int _{S^2} d\hat{\mathbf{\Omega}} \int_{-\infty}^{\infty}df\, 
\tilde{h}_A (f, \hat{\mathbf{\Omega}})\, \nonumber \\
&&e^{2\pi if (t-\hat{\mathbf{\Omega}} \cdot \vec
{\mathbf{X}}/c)} F_A(f, \hat{\mathbf{\Omega}})\:.  \label{eq7}
\end{eqnarray}
Thus, the Fourier transform of Eq.\,(\ref{eq7}) is
\begin{equation}
\tilde{h}(f)=\sum_A \int_{S^2}d\hat{\mathbf{\Omega}}\, \tilde{h}_A(f,\hat{\mathbf
{\Omega}} )e^{-2\pi if \hat{\mathbf{\Omega}} \cdot \vec{\mathbf{X}}/c} F_A (f, \hat{\mathbf
{\Omega}})\:.
\label{eq11}
\end{equation}
We assume that a stochastic GWB is (i) isotropic, (ii) independently polarized (not correlated between polarizations) \footnote{In the theories in five dimensions, the breathing and longitudinal modes are correlated, since their degrees of freedom come from a single polarization mode on 5-dimensional spacetime.}, (iii) stationary, and (iv) Gaussian, which are discussed in \cite{bib7}. Then,
all the statistical properties of the GWB are characterized by
\begin{equation}
\langle \tilde{h}_A^{\ast} (f,\hat{\Omega} ) \tilde{h}_{A^{\prime}} (f^{\prime},\hat{\Omega^
{\prime}} ) \rangle = \delta(f-f^{\prime})\frac{1}{4\pi} \delta^2 ( \hat{\Omega},\hat
{\Omega^{\prime}}) \delta_{AA^\prime} \frac{1}{2} S_h^A (|f|)\;,
\label{eq20}
\end{equation}
where $\delta^2 ( \hat{\Omega},\hat{\Omega^{\prime}}) \equiv \delta(\phi-\phi^{\prime}) 
\delta (\cos\theta - \cos\theta^\prime )$, and $\langle \cdots \rangle$ denotes ensemble average. $S_h^A(f)$ is the one-sided power spectral density of each polarization mode. 

Conventionally, the amplitude of GWB for each polarization is characterized by an energy density per logarithmic frequency bin, normalized by the critical energy density of the Universe:
\begin{equation}
\Omega_{\rm{gw}}^A (f) \equiv \frac{1}{\rho_c}\frac{d\rho_{\rm{gw}}^A}{d \ln f}\:,
\label{eq21}
\end{equation}
where $\rho_c = 3 H_0^2/8\pi G$ and $H_0$ is the Hubble constant. $\Omega_{\rm{gw}}(f)$ is related to $S_h(f)$ by \cite{bib7,bib8}
\begin{equation}
\Omega_{\rm{gw}}^A (f) = \left( \frac{2 \pi^2}{3H_0^2} \right) f^3 S_h^A (f)\:.
\label{eq22}
\end{equation}
Note that the above definition is different from that in the literature \cite{bib7,bib8}, by a factor of 2, since it is defined for each polarization.
It is convenient to represent the energy density with the form $h_{0}^2 \, \Omega_{\rm{gw}}(f)$ by parametrizing the Hubble constant as $H_0=100\,h_{0}\,\rm{km\, s^{-1}\,Mpc^{-1} }$. We assume that $+$ and $\times$ modes are not polarized (The detectability of circular polarizations in the polarized case has been discussed in \cite{bib31,bib32,bib33,bib17}.). We also assume that $x$ and $y$ modes are not polarized. In most of the cosmological scenarios, these assumptions are valid. Then, the GWB energy density of tensor, vector, and scalar modes can be written as
\begin{eqnarray}
\Omega_{\rm{gw}}^T &\equiv & \Omega_{\rm{gw}}^{+} + \Omega_{\rm{gw}}^{\times}  \quad \quad  (\Omega_{\rm{gw}}^{+} = \Omega_{\rm{gw}}^{\times} )\;, 
\label{eq23} \\
\Omega_{\rm{gw}}^V & \equiv& \Omega_{\rm{gw}}^x + \Omega_{\rm{gw}}^y \quad \quad  (\Omega_{\rm{gw}}^{x} = \Omega_{\rm{gw}}^{y} )\;, \\
\Omega_{\rm{gw}}^S &\equiv & \Omega_{\rm{gw}}^b + \Omega_{\rm{gw}}^{\ell} \nonumber \\
&=& \Omega_{\rm{gw}}^b (1+\kappa) \;, 
\label{eq24}
\end{eqnarray}
where the ratio of the energy density in the longitudinal mode to that in the breathing mode is characterized by the parameter $\kappa \equiv \Omega_{\rm{gw}}^{\ell}/\Omega_{\rm{gw}}^{b}$. 

Let us consider the outputs of a detector, $s(t)=h(t)+n(t)$, where $h(t)$ and $n(t)$ are the GW signal and the noise of a detector. We assume that the amplitude of GWB is much smaller than detector noise. Cross-correlation signal $Y$ between two detectors is defined as
\begin{equation}
Y \equiv \int_{-T/2}^{T/2} dt \int_{-T/2}^{T/2} dt^{\prime}\, s_I(t) s_J(t^{\prime} ) Q(t-t^
{\prime})\;,
\label{eq8}
\end{equation}
where $s_I$ and $s_J$ are outputs from the $I$-th and $J$-th detectors, and $T$ is observation time. $Q(t-t^{\prime})$ is a filter function, which is later introduced so that signal-to-noise ratio (SNR) is maximized. In terms of the Fourier representation, we obtain
\begin{equation}
Y=\int_{-\infty}^{\infty}df \int_{-\infty}^{\infty}df^{\prime} \delta _T (f-f ^{\prime}) \tilde{s}_I^
{*}(f) \tilde{s}_J(f^{\prime}) \tilde{Q}(f^{\prime}),
\label{eq9}
\end{equation}
where $\tilde{s}_1(f)$, $\tilde{s}_2(f)$ and $\tilde{Q}(f)$ are the Fourier transforms of 
$s_1(t)$, $s_2(t)$ and $Q(t-t^{\prime})$, respectively. The function $\delta_T(f)$ is defined by
\begin{equation}
\delta _T (f)\equiv \int_{-T/2}^{T/2} dt \,e^{-2\pi ift}=\frac{\sin(\pi fT)}{\pi f}\:.
\nonumber
\end{equation}
In the above derivation, we took the limit of large $T$ for one of the integrals. This is justified by the fact that, in general, $Q(t-t^{\prime})$ rapidly decreases for large $|t-t^{\prime}|$. 
In the absence of intrinsic noise correlation, the correlation signal obtained above ideally has a contribution from only the GWs. Thus, taking the ensemble average of Eq.\,(\ref{eq9}) leads to
\begin{eqnarray}
\mu &\equiv& \langle Y \rangle \nonumber \\
&=&\int_{-\infty}^{\infty}df \int_{-\infty}^{\infty}df^{\prime} \delta_T(f-f^{\prime}) \langle \tilde{h}_I^{*}(f) \tilde{h}_J(f^{\prime}) 
\rangle \tilde{Q}(f^{\prime})\:. \nonumber \\
&& \label{eq10}
\end{eqnarray}
Substituting Eq.\,(\ref{eq11}) into Eq.\,(\ref{eq10}) and using Eqs.\,(\ref{eq20}) and (\ref{eq22}), we obtain
\begin{eqnarray}
\mu &=& \frac{3H_0^2}{4\pi^2} T \int_{-\infty}^{\infty} df \sum_{A} |f|^{-3} \Omega_{\rm{gw}}^A (f) \nonumber \\
&&\times \left[ \int_{S^2} \frac{d \hat{\mathbf{\Omega}}}{4\pi} F_I^A F_J^A e^{2\pi if \hat{\mathbf{\Omega}}\cdot \Delta \vec{\mathbf{X}}/c } \right] \tilde{Q}(f) \nonumber \\
&=& \frac{3H_0^2}{20\pi^2} T \int_{-\infty}^{\infty} df |f|^{-3} \nonumber \\
&& \times \biggl[ \Omega_{\rm{gw}}^T \gamma^T + \Omega_{\rm{gw}}^V \gamma^V +\xi \, \Omega_{\rm{gw}}^S \gamma^S \biggr] \tilde{Q}(f) \;, 
\label{eq30}
\end{eqnarray}
where we defined
\begin{equation}
\Delta \vec{\mathbf{X}} \equiv \vec{\mathbf{X}}
_I-\vec{\mathbf{X}}_J \;, \nonumber 
\end{equation}
and 
\begin{equation}
\xi \equiv \frac{1}{3} \biggl( \frac{1+2\kappa}{1+\kappa} \biggr) \;. \nonumber 
\end{equation}
The parameter $\xi$ is in the range $1/3 \leq \xi \leq 2/3$ and characterizes the ratio of the energy in the longitudinal mode to the breathing mode. We also defined overlap reduction functions 
\begin{eqnarray}
\gamma_{IJ}^{T}(f) &\equiv& \frac{5}{2} \int_{S^2} \frac{d\hat{\mathbf{\Omega}}}{4 \pi}\, e^{2\pi i f \hat{\Omega}\cdot \Delta \vec{X}/c} (F_I^{+} F_J^{+} +F_I^{\times} F_J^{\times}) \;, \nonumber \\
&& \label{eq13} \\
\gamma_{IJ}^{V}(f) &\equiv& \frac{5}{2} \int_{S^2} \frac{d\hat{\mathbf{\Omega}}}{4 \pi}\, e^{2\pi i f \hat{\Omega}\cdot \Delta \vec{X}/c} (F_I^{x} F_J^{x} +F_I^{y} F_J^{y}) \;, \nonumber \\
&& \label{eq28} \\
\gamma_{IJ}^{S}(f) &\equiv& \frac{15}{1+2\kappa} \int_{S^2} \frac{d\hat{\mathbf{\Omega}}}{4 \pi}\, e^{2\pi i f \hat{\Omega}\cdot \Delta \vec{X}/c} (F_I^{b} F_J^{b} +\kappa F_I^{\ell} F_J^{\ell}) \;. \nonumber \\
&& \label{eq29}
\end{eqnarray}
These functions are normalized so that they give unity in the low-frequency limit, which is easily verified by using Eqs.\,(\ref{eq4}) - (\ref{eq5}). The overlap reduction function represents how much degree of correlation between detectors in the GW signal is preserved.

Next, we will calculate the variance of the correlation signal. Here we assume that noises in two detectors do not correlate at all and that the magnitude of the GW signal is much smaller than that of noise. Consequently, the variance of the correlation signal is
\begin{equation}
\sigma ^2 \equiv  \langle Y^2 \rangle - \langle Y \rangle ^2 \approx \langle Y^2 \rangle\:,
\label{eq14}
\end{equation}
where the weak-signal assumption for the GWB is used. Then, using Eq.\,(\ref{eq9}), it follows
\begin{eqnarray}
\sigma ^2&\approx& \int_{-\infty}^{\infty} df \int_{-\infty}^{\infty} df^{\prime}\, \tilde{Q}(f) \tilde{Q}(f^{\prime})\, \nonumber \\
&& \times \langle 
\tilde{s}^{\ast}_I(f) \tilde{s}_J(f) \tilde{s}_I^{\ast}(f^{\prime}) \tilde{s}_J
(f^{\prime}) \rangle \nonumber \\
&\approx& \int_{-\infty}^{\infty} df \int_{-\infty}^{\infty} df^{\prime}\, \tilde{Q}(f) \tilde{Q}^{\ast}(f^{\prime})\, \nonumber \\
&& \times \langle 
\tilde{n}^{\ast}_I(f) \tilde{n}_I(f^{\prime}) \rangle \, \langle \tilde{n}_J(f) \tilde{n}^{\ast}_J
(f^{\prime}) \rangle \nonumber \\
&\approx &\frac{T}{4} \int_{-\infty}^{\infty} df \,P_I(|f|) P_J(|f|)\, | \tilde{Q}(f) | ^2\:, \label{eq15}
\end{eqnarray}
where the one-sided power spectrum density of noise is defined by
 \begin{equation}
\langle \tilde{n}^{\ast}_I(f) \tilde{n}_I(f^{\prime}) \rangle \equiv \frac{1}{2}\delta (f-f^
{\prime})P_I(|f|) \nonumber, \;\;\;\;\;\;\;\; i=1,2\:.
\end{equation}

Now we can determine the form of the optimal filter $\tilde{Q}(f)$. Equations (\ref{eq30}) 
and (\ref{eq15}) are expressed more simply, using an inner product
\begin{equation}
(A, B) \equiv \int_{-\infty}^{\infty} df A^{\ast}(f) B(f) P_I(|f|) P_J(|f|)\:,
\nonumber
\end{equation} 
as
\begin{eqnarray}
\mu &=& \frac{3H_0^2}{20\pi^2}T \left( \tilde{Q}, \frac{\gamma (|f|) \Omega_{\rm{gw}}(|f|)}{|
f|^3 P_I(|f|) P_J(|f|)} \right)\:,
\label{eq16} \\
\sigma ^2 &\approx & \frac{T}{4} \left( \tilde{Q}, \tilde{Q} \right)\:,
\label{eq17} 
\end{eqnarray}
where $\Omega_{\rm{gw}}^T \gamma^T + \Omega_{\rm{gw}}^V \gamma^V +\xi \, \Omega_{\rm{gw}}^S \gamma^S$ is simply written by $\gamma \Omega_{\rm{gw}}$.
From Eqs.\,(\ref{eq16}) and (\ref{eq17}), the SNR for GWB is defined as
${\rm{SNR}} \equiv \mu / \sigma $. Therefore, the filter function, which maximize the SNR , turns out to be
\begin{equation}
\tilde{Q}(f)=K \,\frac{\gamma (f) \Omega_{\rm{gw}}(|f|)}{|f|^3 P_I(|f|) P_J(|f|)},
\label{eq19}
\end{equation} 
with an arbitrary normalization factor $K$. Applying this optimal filter to the above 
equations, we obtain the optimal SNR
\begin{equation}
{\rm{SNR}} = \frac{3H_0^2}{10\pi^2 } \sqrt{T} \left[ \int_{-\infty}^{\infty} df \frac{\gamma ^2 (|f|) 
\Omega ^2_{\rm{gw}}(|f|)}{f^6 P_I(|f|) P_J(|f|)} \right]
^{1/2}\;, 
\label{eq18}
\end{equation}
where $\Omega_{\rm{gw}} \gamma = \Omega_{\rm{gw}}^T \gamma^T + \Omega_{\rm{gw}}^V \gamma^V +\xi \, \Omega_{\rm{gw}}^S \gamma^S$.

\begin{figure}[t]
\begin{center}
\includegraphics[width=6cm]{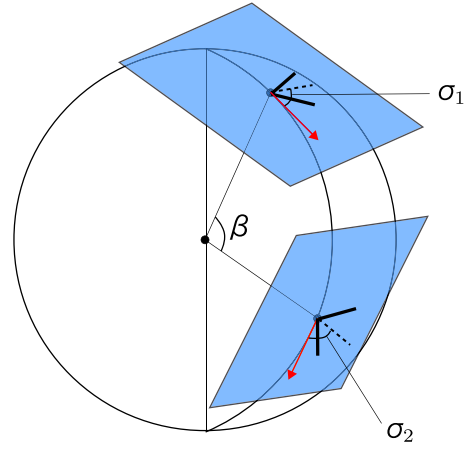}
\caption{(color online). Coordinate system on the Earth for a detector pair.}
\label{fig5}
\end{center}
\end{figure}

\subsection{Overlap reduction function}
Here we present the analytical expressions for the overlap reduction functions given in Eqs.\,(\ref{eq13}) - (\ref{eq29}), performing the angular integrals by expanding the overlap reduction functions with tensorial bases. Following \cite{bib7}, they are, in general, expressed in terms of the detector tensor $D_{ij}$, and a unit vector $\hat{d}_i$, defined by $\hat{d}_i \equiv \Delta \vec{X}/ |\Delta \vec{X}|$:
\begin{eqnarray}
\gamma_{IJ}^{M} (f) &=& \rho_1^{M} (\alpha) D_I^{ij} D^J_{ij} + \rho_2^{M} (\alpha) D_{I,\,k}^{i} D_J^{kj} \hat{d}_i \hat{d}_j \nonumber \\
&&+\rho_3^{M} (\alpha) D_I^{ij} D_J^{k\ell} \hat{d}_i \hat{d}_j \hat{d}_k \hat{d}_{\ell}
\;, \label{eq31}
\end{eqnarray}  
$M=T,V,$ and $S$, together with 
\begin{equation}
\left(
\begin{array}{c} 
\rho_1^{T} \\
\rho_2^{T} \\
\rho_3^{T} 
\end{array}
\right)
=\frac{1}{14}
\left(
\begin{array}{ccc} 
28 & -40 & 2  \\
0 & 120 & -20 \\
0 & 0 & 35
\end{array}
\right)
\left(
\begin{array}{c} 
j_0 \\
j_2 \\
j_4  
\end{array}
\right)\;, \nonumber
\end{equation}
for the tensor mode,
\begin{equation}
\left(
\begin{array}{c} 
\rho_1^{V} \\
\rho_2^{V} \\
\rho_3^{V} 
\end{array}
\right)
=\frac{2}{7}
\left(
\begin{array}{ccc} 
7 & 5 & -2  \\
0 & -15 & 20 \\
0 & 0 & -35
\end{array}
\right)
\left(
\begin{array}{c} 
j_0 \\
j_2 \\
j_4  
\end{array}
\right)\;, \nonumber
\end{equation}
for the vector mode, and 
\begin{equation}
\left(
\begin{array}{c} 
\rho_1^{S} \\
\rho_2^{S} \\
\rho_3^{S} 
\end{array}
\right)
=\frac{1}{7}
\left(
\begin{array}{ccc} 
14 & 20 & 6 \\
0 & -60 & -60 \\
0 & 0 & 105
\end{array}
\right)
\left(
\begin{array}{c} 
j_0 \\
j_2 \\
j_4  
\end{array}
\right)\;, 
\label{eq169}
\end{equation}
for the scalar mode. Details of the derivation are summarized in Appendix \ref{secA}.
Here, $j_n (\alpha)$ is the spherical Bessel function with its argument given by
\begin{equation}
\alpha(f) \equiv \frac{2 \pi f |\Delta \vec{X}|}{c}\;. \nonumber
\end{equation}

To further investigate the dependence of the overlap reduction function on the detector configurations, especially for ground-based detectors, we introduce the coordinate system on the Earth shown in Fig.\,\ref{fig5}. This coordinate system significantly simplifies the expression in Eq.\,(\ref{eq31}). The relative location and orientation of two detectors are characterized by the three parameters, $(\beta,\;\sigma_1,\;\sigma_2)$. The $\beta$ is the separation angle between two detectors, measured from the center of the Earth. The angles $\sigma_1$ and $\sigma_2$, are the orientations of the bisector of two arms of each detector, measured in a counterclockwise manner relative to the great circle connecting the two detectors. The distance between two detectors is given by
\begin{equation}
|\Delta \mathbf{X}| = 2 R_E \,\sin\frac{\beta}{2}\;, \nonumber
\end{equation}
where the radius of the Earth is $R_E= 6371 \,\rm{km}$. Defining new parameters, 
\begin{equation}
\sigma_{+}\equiv \frac{\sigma_1+\sigma_2}{2}\;,\;\;\;\;\;\;\;\; \sigma_{-} \equiv \frac{\sigma_1-\sigma_2}{2}\;, \nonumber
\end{equation}
we can completely fix the relative position and orientation of a detector pair by the three parameters, $(\beta,\;\sigma_{+},\;\sigma_{-})$. In the coordinate system on the Earth, the overlap reduction functions, Eq.\,(\ref{eq31}), can be reduced to
\begin{itemize}
\item{Tensor mode}\\
\begin{eqnarray}
\gamma^T (\alpha,\, \beta,\,\sigma_{+},\,\sigma_{-} ) &=& \Theta _{T+}(\alpha,\, \beta)\,\cos(4\sigma_{+}) \nonumber \\
&&+ \Theta _{T-}(\alpha,\, \beta)\,\cos(4\sigma_{-}) \;,
\label{eq38}
\end{eqnarray}
\begin{eqnarray}
\Theta _{T+} (\alpha,\,\beta) &\equiv & - \left( \frac{3}{8} j_0 -\frac{45}{56} j_2 + \frac{169}{896} j_4 \right) \nonumber \\
&&+ \left( \frac{1}{2} j_0 -\frac{5}{7} j_2 - \frac{27}{224} j_4 \right) \cos \beta \nonumber \\ 
&&- \left( \frac{1}{8} j_0 +\frac{5}{56} j_2 + \frac{3}{896} j_4 \right) \cos 2\beta \;, \nonumber \\
&& \label{eq32} \\
\Theta _{T-} (\alpha,\,\beta) &\equiv & \left( j_0 +\frac{5}{7} j_2 + \frac{3}{112} j_4 \right) \cos \left( \frac{\beta}{2} \right)^4 \;,
\label{eq33}
\end{eqnarray}
\item{Vector mode}\\
\begin{eqnarray}
\gamma^V (\alpha,\,\beta,\,\sigma_{+},\,\sigma_{-})&=& \Theta _{V+}(\alpha,\,\beta)\,\cos(4\sigma_{+}) \nonumber \\
&& + \Theta _{V-}(\alpha,\,\beta)\,\cos(4\sigma_{-}) \;, 
\label{eq39}
\end{eqnarray}
\begin{eqnarray}
\Theta _{V+} (\alpha,\,\beta) &\equiv & - \left( \frac{3}{8} j_0 +\frac{45}{112} j_2 - \frac{169}{224} j_4 \right) \nonumber \\
&&+ \left( \frac{1}{2} j_0 +\frac{5}{14} j_2 + \frac{27}{56} j_4 \right) \cos \beta \nonumber \\ 
&&- \left( \frac{1}{8} j_0 -\frac{5}{112} j_2 - \frac{3}{224} j_4 \right) \cos 2\beta \;, \nonumber \\
&& \label{eq34} \\
\Theta _{V-} (\alpha,\,\beta) &\equiv & \left( j_0 -\frac{5}{14} j_2 - \frac{3}{28} j_4 \right) \cos \left( \frac{\beta}{2} \right)^4 \;,
\label{eq35}
\end{eqnarray}
\item{Scalar mode}\\
\begin{eqnarray}
\gamma^S (\alpha,\,\beta,\,\sigma_{+},\,\sigma_{-})&=& \Theta _{S+}(\alpha,\,\beta)\,\cos(4\sigma_{+}) \nonumber \\
&&+ \Theta _{S-}(\alpha,\,\beta)\,\cos(4\sigma_{-}) \;, 
\label{eq40}
\end{eqnarray}
\begin{eqnarray}
\Theta _{S+} (\alpha,\,\beta) &\equiv & - \left( \frac{3}{8} j_0 +\frac{45}{56} j_2 + \frac{507}{448} j_4 \right) \nonumber \\
&&+ \left( \frac{1}{2} j_0 +\frac{5}{7} j_2 - \frac{81}{112} j_4 \right) \cos \beta \nonumber \\ 
&&- \left( \frac{1}{8} j_0 -\frac{5}{56} j_2 + \frac{9}{448} j_4 \right) \cos 2\beta \;, \nonumber \\
&& \label{eq36} \\
\Theta _{S-} (\alpha,\,\beta) &\equiv & \left( j_0 -\frac{5}{7} j_2 + \frac{9}{56} j_4 \right) \cos \left( \frac{\beta}{2} \right)^4 \;.
\label{eq37}
\end{eqnarray}
\end{itemize}

\begin{table}[t]
\begin{center}
\begin{tabular}{|l|c|c|c|}
\hline
interferometer & $\theta$ & $\phi$ & $\psi$ \\
\hline
AIGO (A) & 121.4 & 115.7 & -45.0 \\ 
LCGT (C) & 53.6 & 137.3 & 70.0 \\
LIGO-H1 (H) & 43.5 & -119.4 & 171.8 \\
LIGO-L1 (L) & 59.4 & -90.8 & 243.0 \\
VIRGO (V) & 46.4 & 10.5 & 116.5 \\
TAMA300 (T) & 54.3 & 139.5 & 225.0 \\
GEO600 (G) & 47.7 & 9.8 & 68.8 \\
\hline
\end{tabular}
\end{center}
\caption{Positions and orientations of kilometer-sized interferometers on the Earth (in unit of degree), provided in \cite{bib17}.}
\label{tab4}
\end{table}

\begin{table}[t]
\begin{center}
\begin{tabular}{|l|c|c|c|c|c|}
\hline
detector pair & $\beta$ & $\sigma_{+}$ & $\sigma_{-}$ & separation [km] & $f_c$ [Hz] \\
\hline
A - C & 70.8 & 31.4 & 31.9 & 7.38 $\times 10^3$ & 20 \\ 
A - H & 135.6 & 45.1 & 53.7 & 1.18 $\times 10^4$ & 13 \\
A - L & 157.3 & 2.1 & 38.0 & 1.25 $\times 10^4$ & 12 \\
A - V & 121.4 & 60.8 & 20.2 & 1.11 $\times 10^4$ & 13 \\
C - H & 72.4 & 25.6 & 89.1 & 7.52 $\times 10^3$ & 20 \\
C - L & 99.2 & 68.1 & 42.4 & 9.71 $\times 10^3$ & 15 \\
C - V & 86.6 & 5.6 & 28.9 & 8.74 $\times 10^3$ & 17 \\
H - L & 27.2 & 62.2 & 45.3 & 3.00 $\times 10^3$ & 51 \\
H - V & 79.6 & 55.1 & 61.1 & 8.16 $\times 10^3$ & 18 \\
L - V & 76.8 & 83.1 & 26.7 & 7.91 $\times 10^3$ & 19 \\
\hline
\end{tabular}
\end{center}
\caption{Relative positions and orientations of detector pairs on the Earth (in unit of degree), and separation between two detectors and the characteristic frequency of the overlap reduction function. Each detector is represented by initial letters indicated in Table \ref{tab4}.}
\label{tab2}
\end{table}

\begin{figure*}[t]
\begin{center}
\includegraphics[width=15cm]{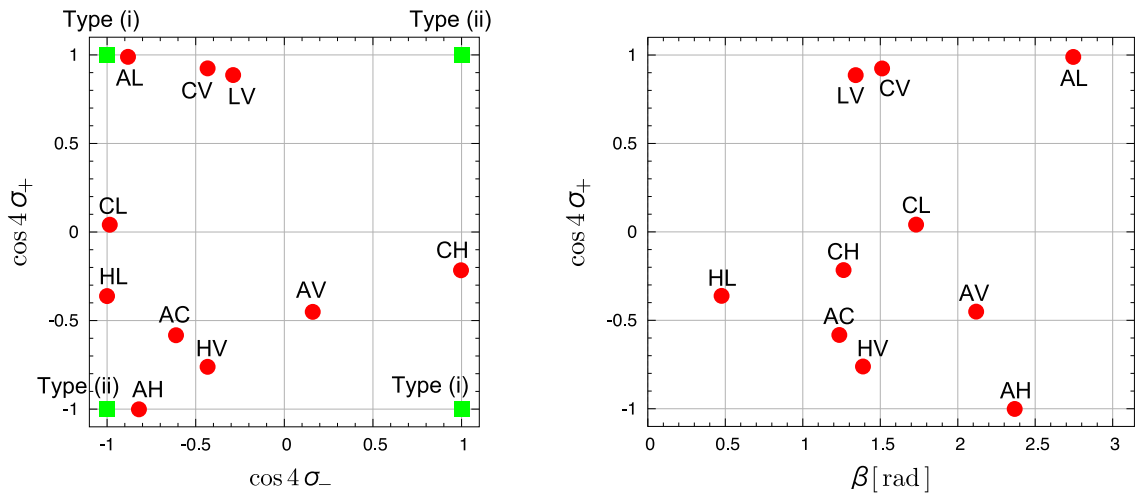}
\caption{(color online). Relative positions and orientations of a specific detector pair. The left panel shows the combinations $(\cos4\sigma_- , \cos4\sigma_+)$. The right panel shows the combinations $(\beta , \cos4\sigma_+)$. The possible optimal configuration, types (i) and (ii), are also shown in the left panel.}
\label{fig11}
\end{center}
\end{figure*}

\begin{figure*}[t]
\begin{center}
\includegraphics[width=16.5cm]{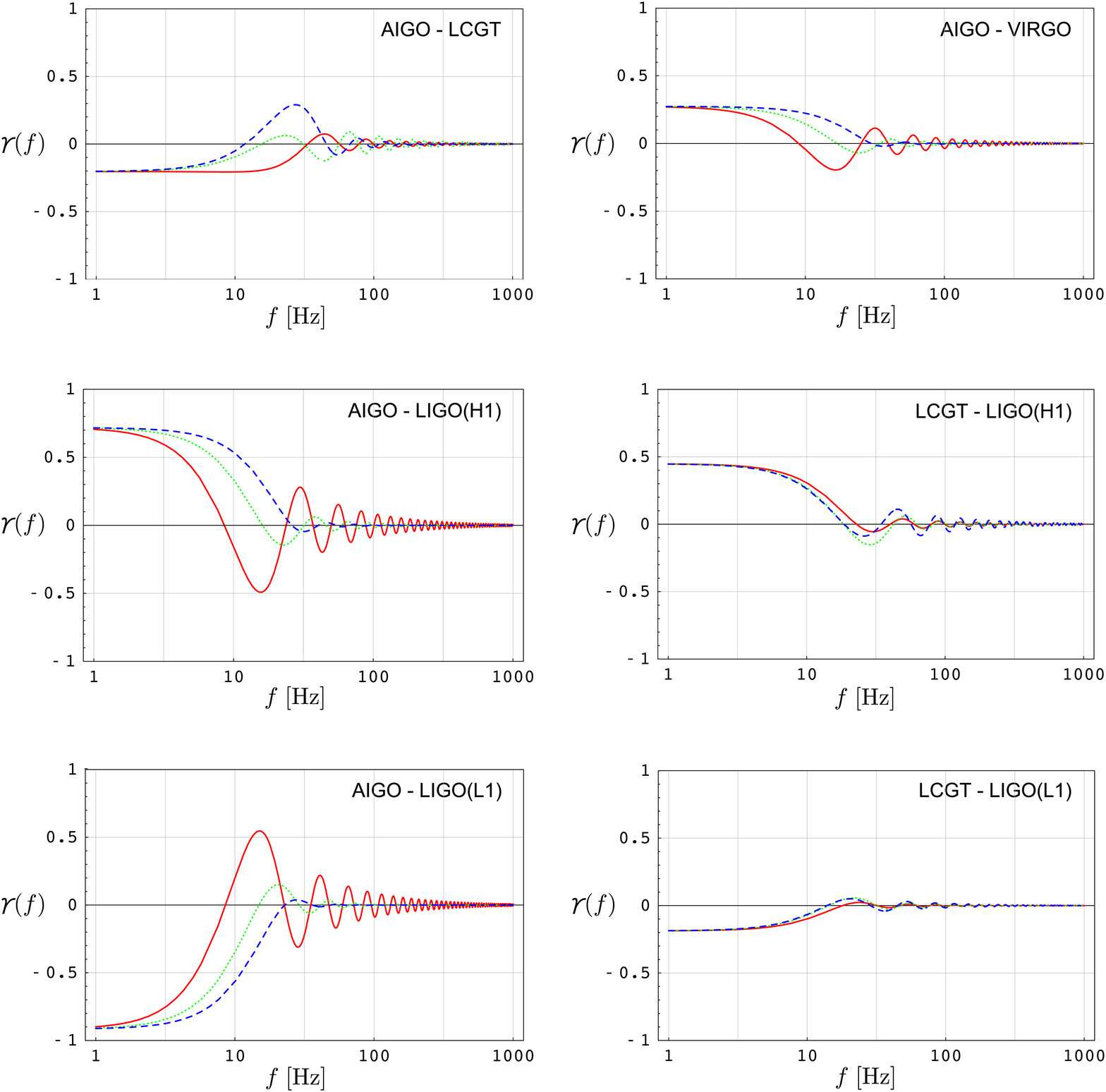}
\caption{(color online). Overlap reduction functions for real-detector pairs on the Earth. Each curve shows tensor mode (red, solid), vector mode (green, dotted), and scalar mode (blue, dashed).}
\label{fig6}
\end{center}
\end{figure*}

\begin{figure*}[t]
\begin{center}
\includegraphics[width=16.5cm]{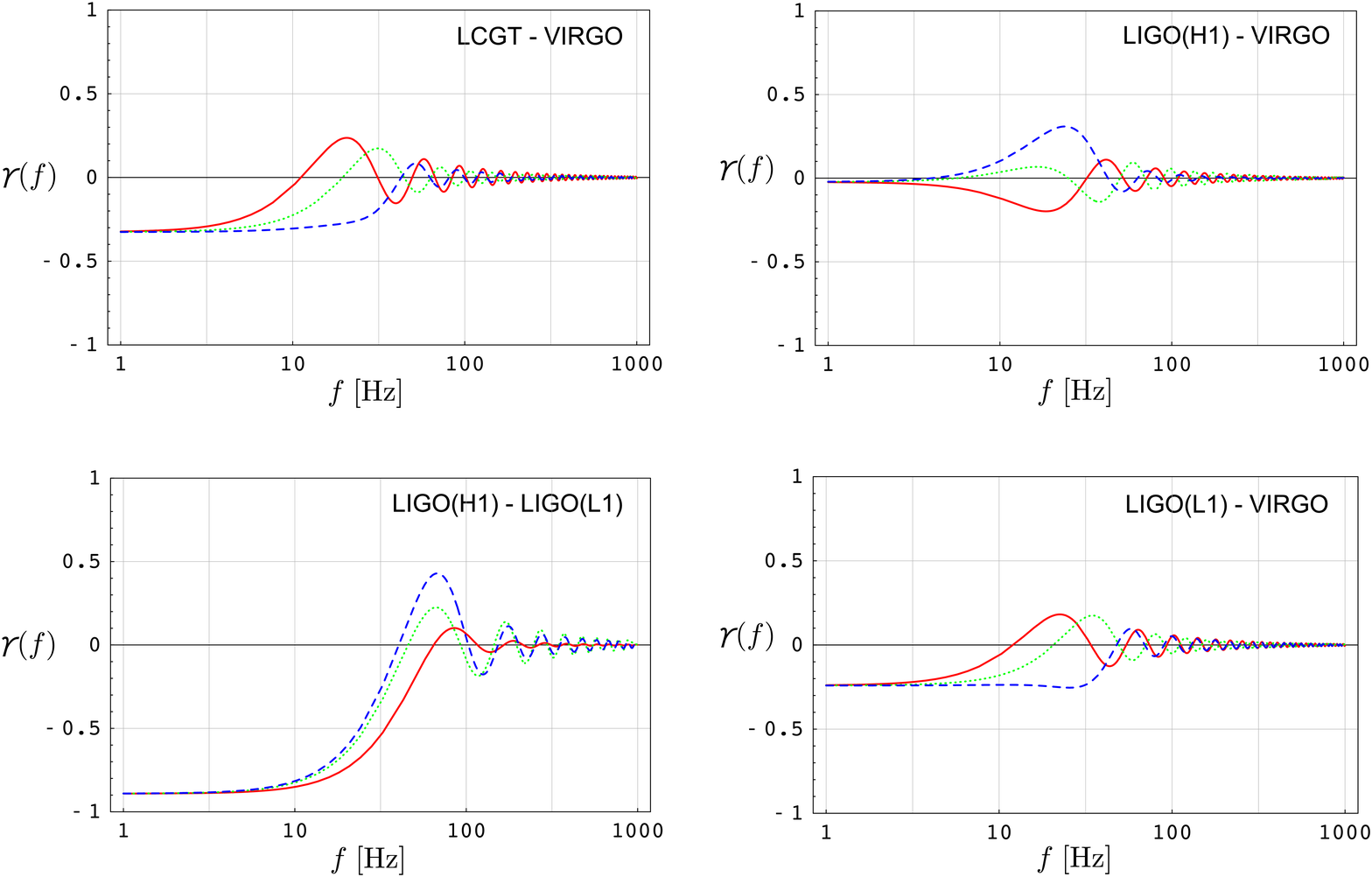}
\caption{(color online). Overlap reduction functions for real-detector pairs on the Earth. Each curve shows tensor mode (red, solid), vector mode (green, dotted), and scalar mode (blue, dashed).}
\label{fig7}
\end{center}
\end{figure*}

\subsection{Overlap reduction functions of specific detectors}
Let us consider the overlap reduction functions for an existing detector pair on the Earth. Given the relative coordinates $(\beta,\;\sigma_{+},\;\sigma_{-})$ of a detector pair, the overlap reduction functions can be plotted as a function of frequency. 

The positions and orientations of the currently operating and planned kilometer-size interferometers are listed in Table \ref{tab4}. To specify the detector positions on the Earth, we use a spherical coordinate system ($\theta,\;\phi$) with which the north pole is at $\theta=0^{\circ}$, and $\phi$ represents the longitude. The orientation angle $\psi$ is the angle between the local east direction and the bisecting line of two arms of each detector measured counterclockwise. Hereafter we will consider only advanced (the second-generation) detectors: AIGO \cite{bib13}, LCGT \cite{bib12}, advanced LIGO (H1) and LIGO (L1) \cite{bib14}, and advanced VIRGO \cite{bib15}, because the pairs of the advanced interferometers are more sensitive to a GWB and have more opportunity to detect a GWB. From the positions and orientations in Table \ref{tab4}, the relative positions and orientations $(\beta,\;\sigma_{+},\;\sigma_{-})$ for each detector pair are determined, and are listed in Table \ref{tab2}. The combinations are also illustrated in Fig.\,\ref{fig11}. The overlap reduction functions, calculated with the parameter set $(\beta,\;\sigma_{+},\;\sigma_{-})$ for a  real-detector pair, are shown in Fig.\,\ref{fig6} and Fig.\,\ref{fig7}. 

The overlap reduction functions start to oscillate and decay rapidly above the characteristic frequency $f_c$, given by $f_c \equiv c/(2|\Delta \mathbf{X}|)$. Numerical values of $f_c$ for specific detector pairs are listed in Table \ref{tab2}. At low frequencies, the functions approach constant values, whose value are determined by the relative orientation of the detector pair. The difference of the behavior between the polarization modes appears at around the characteristic frequency. Mathematically, this is because the coefficients of $j_0$ in Eqs.\,(\ref{eq32}) and (\ref{eq33}), (\ref{eq34}) and (\ref{eq35}), and (\ref{eq36}) and (\ref{eq37}) are exactly the same, while the coefficients of $j_2$ and $j_4$ are different ($j_0\rightarrow1$, $j_2\rightarrow0$, and $j_4\rightarrow0$). At much higher frequencies, since the overlap reduction function significantly reduces, the most interesting frequency range is around the characteristic frequency, e.g., $\sim 10-100\,\rm Hz$ for the detectors on the Earth.

\subsection{Optimal detector configuration}
\begin{figure}[t]
\begin{center}
\includegraphics[width=7cm]{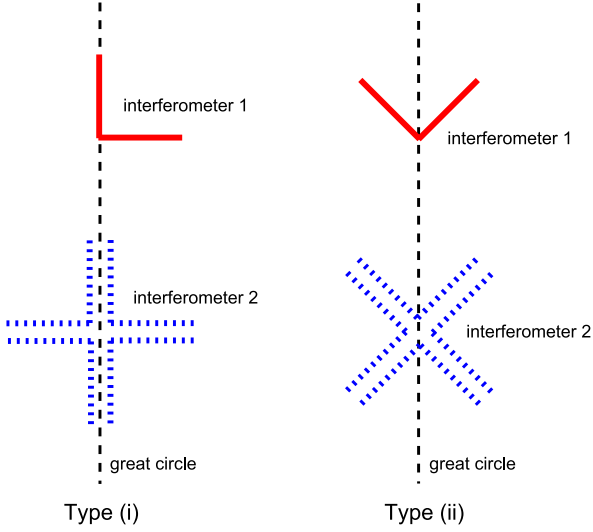}
\caption{(color online). Optimal configurations of a detector pair. For a fixed orientation of the interferometer 1, all possible orientations of the optimal interferometer 2 for a fixed separation $\beta$ are illustrated.}
\label{fig13}
\end{center}
\end{figure}

There are three parameters specifying the detector configuration. We seek an optimal detector configuration, which maximizes the SNR given in Eq.\,(\ref{eq18}).
From the overlap reduction function in Eqs.\,(\ref{eq38}), (\ref{eq39}), and (\ref{eq40}), the optimal configuration of the detectors can be classified into two types:
\begin{eqnarray}
{\rm Type}\; ({\rm{i}}): \; \; &&\cos(4 \sigma_+) = - \cos(4 \sigma_-) = \pm 1 \;, \nonumber \\
{\rm Type}\; ({\rm{ii}}):\;  &&\cos(4 \sigma_+) = \cos(4 \sigma_-) = \pm 1 \;, \nonumber 
\end{eqnarray}
which are illustrated in Fig.\,\ref{fig13}. In type (i), the solutions are $\sigma_1=\pi/4 \mod \pi,\,\sigma_2=-\pi/4 \mod \pi$ for the plus sign, and $\sigma_1=\pi/4 \mod \pi ,\, \sigma_2=\pi/4 \,\mod \pi$ for the minus sign. This means that the great circle connecting two detectors is parallel to one of the arms of both detectors. As for type (ii), the solutions are $\sigma_1=0 \mod \pi,\,\sigma_2=0 \mod \pi$ for the plus sign, and $\sigma_1=\pi/2 \mod \pi ,\, \sigma_2=0 \,\mod \pi$ for the minus sign. This corresponds to the case in which the great circle connecting two detectors is parallel to the bisector of the two arms of each detector, or is directed in the orientations that one of the detectors is rotated by multiples of $\pi/2$ from the former. Therefore, the optimal configuration is realized when one of the arms of the two detectors is parallel or rotated by multiples of 45 degrees, relative to the great circle connecting two detectors. Note that both types of the configurations are not simultaneously optimal one. Whether the configuration is optimal or not depends on the signs of the functions $\Theta _{M+}$ and $\Theta _{M-}$, $M=T,V,S$.

\subsection{SNR}
We calculate the SNR for each mode with two detectors, assuming that only one polarization mode (tensor, vector, or scalar mode) exists. The separation of the polarization modes is addressed in the next section.

The SNR for each polarization mode can be calculated by using the formula, Eq.\,(\ref{eq18}). As for the power spectra of the detector noise $P_I(f)$, we assume that, for simplicity, all advanced detectors (A, C, H, L, V) have the same noise as that of advanced LIGO. The analytical fit of the noise power spectrum of the advanced LIGO, based on \cite{bib16}, is given by \cite{bib17}
\begin{widetext}
\begin{equation}
P(f) = \left\{
\begin{array}{lll} 
\displaystyle
10^{-44} \left( \frac{f}{10\,{\rm Hz}} \right)^{-4} + 10^{-47.25} \left( \frac{f}{100\,{\rm Hz}} \right)^{-1.7} {\rm Hz}^{-1} 
\quad \quad \quad \quad \;\;\; {\rm{for}}\;\; 10\,{\rm Hz} \leq f \leq 240\,{\rm Hz} \;,  \nonumber \\
\displaystyle
10^{-46} \left( \frac{f}{1000\,{\rm Hz}} \right)^{3} {\rm Hz}^{-1} 
\quad \quad \quad \quad \quad \quad \quad \quad \quad \quad \quad \quad \quad \quad \quad \;\; {\rm{for}}\;\; 240\,{\rm Hz} \leq f \leq 3000\,{\rm Hz} \;, \nonumber \\
\displaystyle 
\infty \quad \quad \quad \quad \quad \quad \quad \quad \quad \quad \quad \quad \quad \quad \quad \quad \quad \quad \quad \quad \quad \quad \quad \quad \quad \; {\rm otherwise} \;.
\end{array}
\right. \nonumber
\end{equation}
\end{widetext}
In the SNR calculation, we assume that the $\Omega_{\rm gw}$ is independent of frequency, i.e., $\Omega_{\rm gw}$ is const.. It is useful to write the results for each mode with normalized SNR, defined by
\begin{equation}
{\rm{normalized\; SNR}} = \frac{{\rm{SNR}}(\beta)}{{\rm{SNR}}({\beta=0})} \;, \nonumber 
\end{equation}
where 
\begin{equation}
{\rm{SNR}}({\beta=0}) \approx 9.7 \times \left( \frac{T}{3\,\rm{yr}} \right) ^{1/2} \left( \frac{h_0^2 \Omega_{\rm{gw}}}{10^{-9}}\right)  \;, \nonumber 
\end{equation}
($h_0^2 \Omega_{\rm{gw}}$ has to be replaced with $\xi h_0^2 \Omega_{\rm{gw}}$ for the scalar mode). Note that the above value of ${\rm{SNR}}({\beta=0})$ is identical for each polarization mode, because of the degeneracy of them at the low frequencies.

\begin{figure}[t]
\begin{center}
\includegraphics[width=8cm]{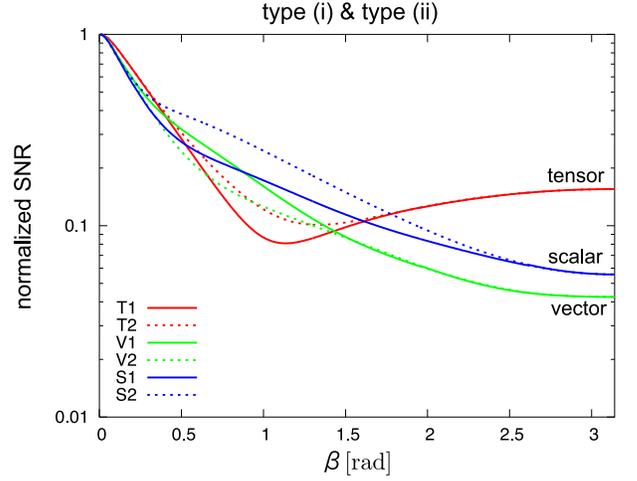}
\caption{(color online). Normalized SNRs of the detector pair of type (i) (solid lines) and type (ii) (dotted lines) as a function of $\beta$. The tensor, vector, and scalar modes are represented by red, green, and blue curves, respectively, as indicated on the right side of the figure.}
\label{fig8}
\end{center}
\end{figure}

\begin{figure}[t]
\begin{center}
\includegraphics[width=8cm]{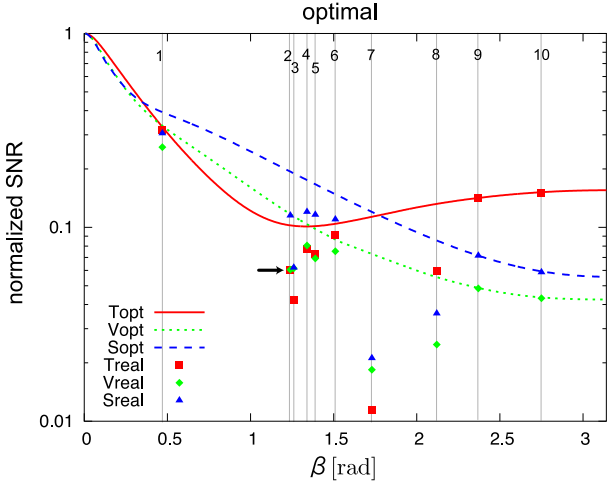}
\caption{(color online). Normalized SNRs of the optimal and real detector pairs as a function of $\beta$. Each curve shows $\max \,\bigl\{ {\rm SNR}|_{\rm type(i)},\,{\rm SNR}|_{\rm type(ii)} \bigr\}$ for the tensor mode (red, solid), the vector mode (green, dotted), and the scalar mode (blue, dashed). The squares (red), diamonds (green), and triangles (blue) are the normalized SNRs of real detector pairs for the tensor, vector, and scalar modes, respectively. At the arrow, four points are overlapped: (detector pair, mode) = (2, tensor), (2, vector), (3, vector), and (3, scalar). The numbers in the figure represent the real-detector pair: 1=HL, 2=AC, 3=CH, 4=LV, 5=HV, 6=CV, 7=CL, 8=AV, 9=AH, and 10=AL.}
\label{fig9}
\end{center}
\end{figure}

In Fig.\,\ref{fig8}, the normalized SNRs for the possible optimal configuration of detector pairs, type (i) and (ii), are shown as a function of $\beta$. Interestingly, most of the detector pairs have almost the same sensitivity to the three modes. In Fig.\,\ref{fig9}, the optimal SNR, $\max \,\bigl\{ {\rm SNR}|_{\rm type(i)},\,{\rm SNR}|_{\rm type(ii)} \bigr\}$, is shown as a function of $\beta$, together with SNRs of specific detector pairs for each mode. The SNRs of the specific detector pairs, except for AIGO-LIGO (H1) and AIGO-LIGO (L1) pairs, are smaller than the optimal one due to the incomplete coincidence of the detector orientations. It is interesting to note that the SNR of the optimal detector configuration for the scalar mode is enhanced at the distance relatively close, while that for the tensor mode is enhanced at the distance relatively far. This feature can be intuitively interpreted by the angular responses of the detector shown in Fig.\,\ref{fig4}. At $\beta \sim \pi/2$, the angular responses of the tensor mode between two detectors are less overlapped than those of the scalar and vector modes. On the other hand, at $\beta \sim \pi$, the angular responses of the tensor mode between two detectors are more overlapped. 

SNRs with a specific detector pair are tabulated in Table \ref{tab5}. As anticipated from Fig.\,\ref{fig9}, a laser-interferometric GW detector is sensitive to a GWB with the non-tensorial polarizations, having almost the same SNR as the tensor mode. 

\begin{table}[b]
\begin{center}
\begin{tabular}{|l|c|c|c|}
\hline
detector pair & $h_0^2 \Omega_{\rm gw}^T $ & $h_0^2 \Omega_{\rm gw}^V$ & $\xi h_0^2 \Omega_{\rm gw}^S$ \\
\hline
A - C & $8.6 \times 10^{-9}$ & $8.6\times 10^{-9}$ & $4.5\times 10^{-9}$ \\ 
A - H & $3.6 \times 10^{-9}$ & $1.1\times 10^{-8}$ & $7.3\times 10^{-9}$ \\
A - L & $3.4 \times 10^{-9}$ & $1.2\times 10^{-8}$ & $8.8\times 10^{-9}$ \\
A - V & $8.7 \times 10^{-9}$ & $2.1\times 10^{-8}$ & $1.4\times 10^{-8}$ \\
C - H & $1.2 \times 10^{-8}$ & $8.4\times 10^{-9}$ & $8.4\times 10^{-9}$ \\
C - L & $4.5 \times 10^{-8}$ & $2.8\times 10^{-8}$ & $2.5\times 10^{-8}$ \\
C - V & $5.7 \times 10^{-9}$ & $6.9\times 10^{-9}$ & $4.7\times 10^{-9}$ \\
H - L & $1.6 \times 10^{-9}$ & $2.0\times 10^{-9}$ & $1.7\times 10^{-9}$ \\
H - V & $7.1 \times 10^{-9}$ & $7.5\times 10^{-9}$ & $4.5\times 10^{-9}$ \\
L - V & $6.7 \times 10^{-9}$ & $6.4\times 10^{-9}$ & $4.3\times 10^{-9}$ \\
\hline
\end{tabular}
\end{center}
\caption{Detectable GWB (${\rm SNR}=5$) with a real-detector pair in the presence of a single polarization mode. The observation time is selected as $T=3\,{\rm yr}$.}
\label{tab5}
\end{table}

\begin{figure}[t]
\begin{center}
\includegraphics[width=8cm]{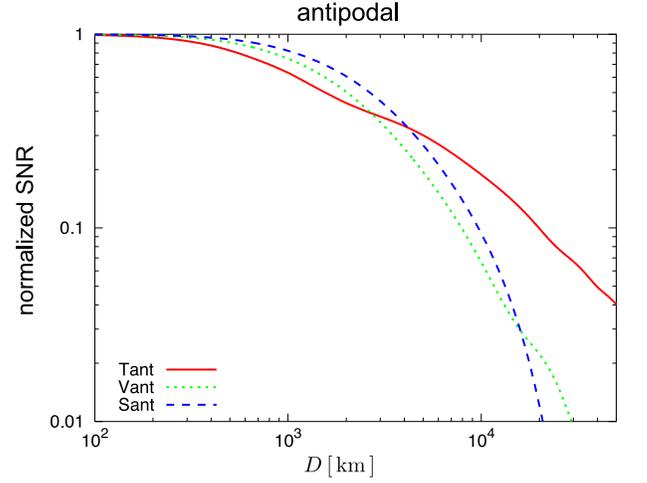}
\caption{(color online). Normalized SNR of an antipodal detector pair as a function of the separation $D$. The tensor, vector, and scalar modes are represented by red (solid), green (dotted), and blue (dashed) curves, respectively. In the case of the (real) Earth, the separation is $D\approx 1.27\times 10^4\,\rm km$.}
\label{fig10}
\end{center}
\end{figure}

For further understanding of the distance dependence of each polarization mode, it is helpful to consider an antipodal detector pair on a virtual earth with an arbitrary radius $R$. The SNRs of the antipodal detector pair depend only on the distance between the detectors. In the antipodal case, setting $\beta=\pi$ and $\sigma_+=\pi/4$ ($\sigma_1=0$ and $\sigma_2=\pi/2$) gives the overlap reduction functions,
\begin{eqnarray}
\gamma_T &=& j_0 - \frac{10}{7} j_2 + \frac{1}{14} j_4 \;, \nonumber \\
\gamma_V &=& j_0 + \frac{5}{7} j_2 - \frac{2}{7} j_4 \;, \nonumber \\
\gamma_S &=& j_0 + \frac{10}{7} j_2 + \frac{3}{7} j_4 \;. \nonumber 
\end{eqnarray}
The separation between the detectors is
\begin{equation}
D \equiv |\Delta \vec{\mathbf{X}}| =2 R \;.
\end{equation}
Given the overlap reduction functions, we can compute normalized SNRs, which are shown in Fig.\,\ref{fig10}. The SNRs monotonically decrease proportional to the separation between the detectors. At relatively small distance, the SNR for each polarization mode is comparable to each other, while at relatively large distance, the SNR of the tensor mode shows slower decay than that of other modes.

\section{Mode separation}
\label{sec5}
In practical observation of a stochastic GWB, three polarization modes of a GWB are mixed in the detector cross-correlation signal. The decomposition of the modes is an important issue. In this section, in the presence of all polarization modes (tensor, vector, and scalar), we discuss how to separately detect each polarization mode of the stochastic GWB.

Let us consider the case in which three detectors are available. From Eq.\,(\ref{eq9}), the cross-correlation signal between the $I$ th and $J$ th detectors is given by
\begin{equation}
\mu_{IJ}= \langle Y_{IJ} \rangle = \int_{-\infty}^{\infty} df \langle \tilde{s}_I^{\ast} (f) \tilde{s}_J (f) \rangle \tilde{Q}(f) \;,
\label{eq50} \\
\end{equation}
Comparing Eq.\,(\ref{eq9}) with Eq.\,(\ref{eq30}), we define a statistic,
\begin{eqnarray}
Z_{IJ} (f) &\equiv& \frac{20\pi^2}{3H_0^2T} |f|^3 \tilde{s}_I^{\ast} (f) \tilde{s}_J (f) \nonumber \\
&=& \Omega_{\rm{gw}}^T(f) \gamma^T_{IJ} (f)+ \Omega_{\rm{gw}}^V(f) \gamma^V_{IJ} (f) \nonumber \\
&&+\xi \, \Omega_{\rm{gw}}^S(f) \gamma^S_{IJ} (f) + {\rm{a \;noise\; term}} \;. \nonumber 
\end{eqnarray}
This statistic contains the contribution from noise terms. However, by taking ensemble average, the noise terms vanish, because we are assuming that the detector noises are independent each other. 
Then, we obtain three cross-correlation signals
\begin{equation}
\left(
\begin{array}{c} 
\langle Z_{12} \rangle \\
\langle Z_{23} \rangle \\
\langle Z_{31} \rangle
\end{array}
\right)
= \mathbf{\Pi} 
\left(
\begin{array}{c} 
\Omega_{\rm{gw}}^T  \\
\Omega_{\rm{gw}}^V  \\
\xi \Omega_{\rm{gw}}^S 
\end{array}
\right) \;,
\label{eq49} 
\end{equation}
with a detector correlation matrix
\begin{equation}
\mathbf{\Pi} \equiv
\left(
\begin{array}{ccc} 
\gamma^T_{12} & \gamma^V_{12} & \gamma^S_{12} \\
\gamma^T_{23} & \gamma^V_{23} & \gamma^S_{23} \\
\gamma^T_{31} & \gamma^V_{31} & \gamma^S_{31}
\end{array}
\right)
\end{equation}
where the subscripts, $I,J=1,2,$ and $3$, discriminate the detectors. Therefore, the modes can be separated by inverting Eq.\,(\ref{eq49}), namely, 
\begin{equation}
\left(
\begin{array}{c} 
\Omega_{\rm{gw}}^T  \\
\Omega_{\rm{gw}}^V  \\
\xi \Omega_{\rm{gw}}^S 
\end{array}
\right)
= \mathbf{\Pi}^{-1} 
\left(
\begin{array}{c} 
\langle Z_{12} \rangle \\
\langle Z_{23} \rangle \\
\langle Z_{31} \rangle
\end{array}
\right) \;.
\nonumber 
\end{equation}
The explicit expression of $\mathbf{\Pi}^{-1}$, which we call a separation matrix, is 
\begin{widetext}
\begin{equation}
\mathbf{\Pi}^{-1}
=\frac{1}{\det \mathbf{\Pi}}
\left(
\begin{array}{ccc} 
\gamma_{23}^V \gamma_{31}^S - \gamma_{23}^S \gamma_{31}^V  \;\;& \gamma_{31}^V \gamma_{12}^S - \gamma_{31}^S \gamma_{12}^V \;\;& \gamma_{12}^V \gamma_{23}^S - \gamma_{12}^S \gamma_{23}^V \\
\gamma_{23}^S \gamma_{31}^T - \gamma_{23}^T \gamma_{31}^S  \;\;& \gamma_{31}^S \gamma_{12}^T - \gamma_{31}^T \gamma_{12}^S \;\;& \gamma_{12}^S \gamma_{23}^T - \gamma_{12}^T \gamma_{23}^S \\
\gamma_{23}^T \gamma_{31}^V - \gamma_{23}^V \gamma_{31}^T \;\;& \gamma_{31}^T \gamma_{12}^V - \gamma_{31}^V \gamma_{12}^T \;\;& \gamma_{12}^T \gamma_{23}^V - \gamma_{12}^V \gamma_{23}^T
\end{array}
\right)
\;, 
\label{eq52}
\end{equation}
with
\begin{equation}
\det \mathbf{\Pi} = \gamma^T_{12} (\gamma_{23}^V \gamma_{31}^S - \gamma_{31}^V \gamma_{23}^S)+\gamma^V_{12} (\gamma_{23}^S \gamma_{31}^T - \gamma_{31}^S \gamma_{23}^T)+\gamma^S_{12} (\gamma_{23}^T \gamma_{31}^V - \gamma_{31}^T \gamma_{23}^V) \;. \nonumber
\end{equation}
\end{widetext}
For the mode separation, the condition, $\det \mathbf{\Pi} \neq 0$, is required and should be checked. As we will verify later, in the case of the specific detectors on the Earth, we can safely perform the integration of the SNR. 

Next, we derive the SNR formula for separate detection of each polarization mode with three detectors. The GW signal and detector noise are given by replacing $\tilde{s}_I^{\ast} (f) \tilde{s}_J (f)$ in Eqs.\,(\ref{eq50}) and (\ref{eq15}) with a certain combination
\begin{equation}
\eta_{M1} (f) \tilde{s}_1^{\ast} (f) \tilde{s}_2 (f) + \eta_{M2} (f) \tilde{s}_2^{\ast} (f) \tilde{s}_3 (f) + \eta_{M3} (f) \tilde{s}_3^{\ast} (f) \tilde{s}_1 (f) \;, \nonumber
\end{equation}
where $\eta_{Mi} (f), \, i=1,2,3,\;M=T,V,S$, are the proper components of the inverse of the separation matrix, $\mathbf{\Pi}^{-1}$, e.\,g., to separate the tensor mode, $\eta_{T1}=(\gamma_{23}^V \gamma_{31}^S - \gamma_{23}^S \gamma_{31}^V)/\det \mathbf{\Pi}$, $\eta_{T2}=(\gamma_{31}^V \gamma_{12}^S - \gamma_{31}^S \gamma_{12}^V)/\det \mathbf{\Pi}$, and $\eta_{T3}=(\gamma_{12}^V \gamma_{23}^S - \gamma_{12}^S \gamma_{23}^V )/\det \mathbf{\Pi}$. Then, the GW signal for each mode, $M=T,V,$ and $S$, is
\begin{eqnarray}
\mu_M &=& \frac{3 H_0^2}{20 \pi^2} T \int_{-\infty} ^{\infty} df |f|^{-3} \,\bigl[ \eta_{M1} (f) \langle Z_{12} (f) \rangle \nonumber \\
&&+ \eta_{M2} (f) \langle Z_{23} (f) \rangle +\eta_{M3} (f) \langle Z_{31} (f) \rangle \bigr] \tilde{Q} (f) \;, \nonumber \\
&& \label{eq41}
\end{eqnarray}
and the detector noise (the variance of $\mu$) is
\begin{eqnarray}
\sigma^2_M &=& \frac{T}{4} \int_{-\infty} ^{\infty} df \, \bigl[ \eta_{M1}^2 (f) P_{1} (|f|) P_{2} (|f|) \nonumber \\
& +& \eta_{M2}^2 (f) P_{2} (|f|) P_{3} (|f|) +\eta_{M3}^2 (f) P_{3} (|f|) P_{1} (|f|) \bigr]\, \nonumber \\
&\times& |\tilde{Q}(f)|^2  \;.  
\label{eq42} 
\end{eqnarray}
Hereafter we omit the subscript $M$ for the simplicity of the expressions. Note that, however, $H_g$, $H_n$, and SNR below are different functions for each polarization mode.
Defining 
\begin{eqnarray}
H_{g} (f) &\equiv & \eta_{1} (f) \langle Z_{12} (f) \rangle + \eta_{2}(f) \langle Z_{23}(f) \rangle \nonumber \\
&&+\eta_{3}(f) \langle Z_{31}(f) \rangle \;,  \label{eq53} \\
&& \nonumber \\
H_{n}(f) &\equiv & \bigl[ \eta_{1}^2(f) P_{1}(|f|) P_{2}(|f|) + \eta_{2}^2(f) P_{2}(|f|) P_{3}(|f|) \nonumber \\
&&+\eta_{3}^2(f) P_{3}(|f|) P_{1}(|f|) \bigr]^{1/2} \label{eq54} \;,
\end{eqnarray}
and an inner product
\begin{equation}
(A,B) \equiv \int_{-\infty} ^{\infty} df A^{\ast} (f) B (f) H_n^2 (f) \;, \nonumber
\end{equation}
we can write Eqs.\,(\ref{eq41}) and (\ref{eq42}) into
\begin{eqnarray}
\mu &=& \frac{3 H_0^2}{20 \pi^2} T \biggl( \tilde{Q}, \frac{H_g}{|f|^3 H_n^2} \biggr) \;, 
\label{eq55} \\
\sigma^2 &=& \frac{T}{4} \biggl( \tilde{Q}, \tilde{Q} \biggr) \;. 
\label{eq56}
\end{eqnarray}
Thus, to optimize the SNR, the optimal filter should be chosen as
\begin{equation}
\tilde{Q} (f) = K \frac{H_g(f)}{|f|^3 H_n^2(f)} \;. \nonumber
\end{equation}
Substituting the specific form of the optimal filter in Eqs.\,(\ref{eq55}) and (\ref{eq56}), one obtain the SNR formula after the mode separation as
\begin{equation}
{\rm{SNR}} = \frac{3 H_0^2}{10 \pi^2} \sqrt{T} \biggl[ \int_{-\infty} ^{\infty} df \frac{H_g^2(f)}{|f|^6 H_n^2(f)} \biggr] ^{1/2} \;.
\label{eq43}
\end{equation}

Using this formula, we calculate the SNRs. For simplicity, we assume that all detectors have the same noise power spectrum, $P (f) = P_1(f)=P_2(f)=P_3(f)$. In this case, $H_g^2/H_n^2$ in the integrand of Eq.\,(\ref{eq43}) can be written down explicitly. Substituting Eqs.\,(\ref{eq53}) and (\ref{eq54}) into $H_g^2/H_n^2$ and using the components of Eq.\,(\ref{eq52}) give 
\begin{widetext}
\begin{itemize}
\item{tensor mode}\\
\begin{equation}
\frac{H_g^2(f)}{H_n^2(f)} = \frac{(\Omega_{\rm{gw}}^T)^2}{P^2(f)} W_T(f) \;, \quad \quad
W_T(f) \equiv \frac{(\det \Pi) ^2}{(\gamma_{23}^V \gamma_{31}^S - \gamma_{23}^S \gamma_{31}^V)^2+(\gamma_{31}^V \gamma_{12}^S - \gamma_{31}^S \gamma_{12}^V)^2+(\gamma_{12}^V \gamma_{23}^S - \gamma_{12}^S \gamma_{23}^V)^2} \;, \nonumber
\end{equation} 
\item{vector mode}\\
\begin{equation}
\frac{H_g^2(f)}{H_n^2(f)} = \frac{(\Omega_{\rm{gw}}^V)^2}{P^2(f)} W_V(f) \;, \quad \quad
W_V(f) \equiv \frac{(\det \Pi) ^2}{(\gamma_{23}^S \gamma_{31}^T - \gamma_{23}^T \gamma_{31}^S)^2+(\gamma_{31}^S \gamma_{12}^T - \gamma_{31}^T \gamma_{12}^S)^2+(\gamma_{12}^S \gamma_{23}^T - \gamma_{12}^T \gamma_{23}^S)^2} \;, \nonumber
\end{equation} 
\item{scalar mode}\\
\begin{equation}
\frac{H_g^2(f)}{H_n^2(f)} = \frac{(\Omega_{\rm{gw}}^S)^2}{P^2(f)} W_S(f) \;, \quad \quad
W_S(f) \equiv \frac{(\det \Pi) ^2}{(\gamma_{23}^T \gamma_{31}^V - \gamma_{23}^V \gamma_{31}^T)^2+(\gamma_{31}^T \gamma_{12}^V - \gamma_{31}^V \gamma_{12}^T)^2+(\gamma_{12}^T \gamma_{23}^V - \gamma_{12}^V \gamma_{23}^T)^2} \;, \nonumber
\end{equation} 
\end{itemize}
\end{widetext}

\begin{figure}[t]
\begin{center}
\includegraphics[width=8.5cm]{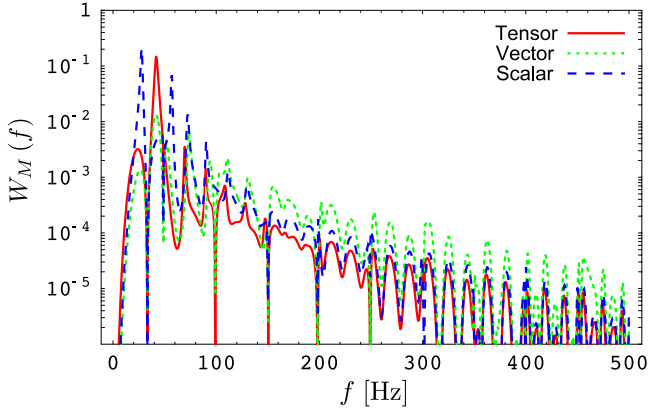}
\caption{(color online). Plot of a squared effective overlap reduction function, $W_M(f)$, for the H-L-V detector set. No divergence of the function occurs in an observational frequency band.}
\label{fig12}
\end{center}
\end{figure}

In the above equations, we define the function $W_M(f)$, $M=T,V,$ and $S$. Comparing Eq.\,(\ref{eq43}) with Eq.\,(\ref{eq18}), we can interpret $\sqrt{W_M(f)}$ as an effective overlap reduction function in the case of the polarization mode separation with three detectors. As we noted earlier, the condition $\det \mathbf{\Pi} \neq 0$ is needed in order to successfully separate the polarization modes. However, we can safely perform the integral, because the contribution to the SNR automatically drops out from the integral when $\det \mathbf{\Pi}= 0$. Another concern about the integral is the pole of $W_M(f)$. For completeness, we checked that $W_M(f)$ of all three-detector sets do not diverge at any frequency in the observation frequency band, as an example, which is shown in Fig.\,\ref{fig12}.

Assuming that GWB spectra are independent of frequency and that observation time is $T=3\,\rm{yr}$, we obtain the SNRs after the mode separation with three specific detectors. The detectable GWB-energy density with advanced detectors with the detection threshold ${\rm{SNR}}=5$ is shown in Table \ref{tab6}. In comparison with the SNRs in the presence of a single polarization mode in Table \ref{tab5}, three detector combinations have almost the same sensitivity to three polarization modes even when the modes are separated.

\begin{table}[t]
\begin{center}
\begin{tabular}{|l|c|c|c|}
\hline
detector set & $h_0^2 \Omega_{\rm gw}^T $ & $h_0^2 \Omega_{\rm gw}^V$ & $\xi h_0^2 \Omega_{\rm gw}^S$ \\
\hline
A - C - H & $5.2 \times 10^{-9}$ & $8.1\times 10^{-9}$ & $5.5\times 10^{-9}$ \\ 
A - C - L & $6.0 \times 10^{-9}$ & $1.5\times 10^{-8}$ & $8.3\times 10^{-9}$ \\
A - C - V & $1.3 \times 10^{-8}$ & $1.0\times 10^{-8}$ & $6.8\times 10^{-9}$ \\
A - H - L & $3.8 \times 10^{-9}$ & $1.2\times 10^{-8}$ & $1.0\times 10^{-8}$ \\
A - H - V & $8.5 \times 10^{-9}$ & $2.2\times 10^{-8}$ & $2.1\times 10^{-8}$ \\
A - L - V & $6.0 \times 10^{-9}$ & $2.4\times 10^{-8}$ & $2.3\times 10^{-8}$ \\
C - H - L & $1.4 \times 10^{-8}$ & $1.9\times 10^{-8}$ & $1.9\times 10^{-8}$ \\
C - H - V & $1.1 \times 10^{-8}$ & $1.0\times 10^{-8}$ & $7.6\times 10^{-9}$ \\
C - L - V & $1.2 \times 10^{-8}$ & $2.0\times 10^{-8}$ & $1.7\times 10^{-8}$ \\
H - L - V & $6.1 \times 10^{-9}$ & $1.3\times 10^{-8}$ & $6.0\times 10^{-9}$ \\
\hline
\end{tabular}
\end{center}
\caption{Detectable energy density of GWB with ${\rm SNR}=5$, assuming $T=3\,{\rm yr}$.}
\label{tab6}
\end{table}

Although we considered the three-detector case above, one can also perform the correlation analysis with more than four detectors \cite{bib7,bib17}. In the case of $N$ detectors available, there are $N_{\rm{pair}} = N(N -1)/2$ correlation signals from detector pairs. Then, the number of independent combinations for the three-mode separation is $N_{\rm{pair}}-2$. The additional combinations can be exploited in order to optimally weight the correlation signals and to enhance the SNR. Such an analysis with multiple-detector pairs has been done in \cite{bib7,bib17}. However, the SNR is not significantly improved when noisy or largely separated detectors are added, because the SNR is mainly determined by the sensitive three detectors among the detector network. Therefore, in the case of more than four detectors, the SNRs are not much different from those provided in Table \ref{tab6}.

\section{Discussions and conclusions}
\label{sec6}
In this paper, we investigated the detectability of additional polarization modes of a stochastic GWB with ground-based laser-interferometric detectors. Such polarization modes, in general, appear in the extended theories of gravitation and the theories with extra dimensions, and can be utilized to constrain the theories beyond GR in a model-independent way. We extended the formalism of the cross-correlation analysis, including the additional polarization modes, and calculated the detectable energy density of the GWB. In the presence of a single polarization mode, a detector pair has almost the same sensitivity to tensor, vector, and scalar modes, and can detect the GWB of $h_0^2 \Omega_{\rm{gw}} \sim 10^{-9}$ with ${\rm{SNR}}=5$. We showed that the mixture of the three polarization modes in the correlation signals can be separated with more than three independent detectors at different sites, and that the separation does not significantly affect the SNR, which is comparable to the SNR obtained in the presence of a single polarization mode. It is interesting that the most sensitive detector set for one polarization mode
does not necessarily coincide with that for other polarization mode: the best detector set is A-H-L for the
tensor polarization and A-C-H for the vector and scalar
polarizations (The noises of all the detectors are assumed to be identical with that of advanced LIGO.).     

We showed that the existence of the non-tensorial polarization modes in a GWB can be probed with more than three detectors. What we know from the observation is the energy density of the GWB for each polarization, which is present in the Universe.  If the GWB is not detected, we obtain the upper limit. To select the correct theories or constrain the theories, we need to compare the GWB spectrum obtained in the observation with that predicted in a specific theory. So, the derivation of the GWB spectrum in a specific theory is the important issue that should be addressed, but we will leave it for future work.

\begin{acknowledgments}
We thank N. Kanda, T. Nakamura, and N. Seto for helpful discussions. We also acknowledge the anonymous referee for useful comments. This work is supported in part by a Grant-in-Aid for Scientific Research from the Japan Society for Promotion of Science (No. 18740132). K. H is supported by the Max Planck Society.
\end{acknowledgments}

\appendix
\section{Tensorial expansion of overlap reduction function}
\label{secA}
In this Appendix, we provide a tensorial expansion of the overlap reduction function, which allows us to perform the angular integral about a GW-propagating direction, without specifying a detector configuration. 

Defining 
\begin{eqnarray}
\Delta \vec{X} &=& \vec{X}_1-\vec{X}_2 \equiv |\Delta \vec{X}| \hat{d}\;, \nonumber \\
\alpha(f) &\equiv & \frac{2 \pi f |\Delta \vec{X}|}{c}\;, \nonumber
\end{eqnarray}
and 
\begin{eqnarray}
\Gamma _{ijk\ell}^T (\alpha, \hat{d}\,) &\equiv& \frac{5}{2} \sum_A \int_{S^2} \frac{d\hat{\Omega}}{4\pi} \, e^{i \alpha \hat{\Omega}\cdot \hat{d}}\, \tilde{e}_{ij}^A (\hat{\Omega}) \tilde{e}_{k\ell}^A (\hat{\Omega})\;, \nonumber \\
&& \label{eq44} \\
\Gamma _{ijk\ell}^V (\alpha, \hat{d}\,) &\equiv& \frac{5}{2} \sum_A \int_{S^2} \frac{d\hat{\Omega}}{4\pi} \, e^{i \alpha \hat{\Omega}\cdot \hat{d}}\, \tilde{e}_{ij}^A (\hat{\Omega}) \tilde{e}_{k\ell}^A (\hat{\Omega})\;, \nonumber \\
&& \label{eq57} \\
\Gamma _{ijk\ell}^S (\alpha, \hat{d}\,) &\equiv& \frac{15}{1+2\kappa} \sum_A \int_{S^2} \frac{d\hat{\Omega}}{4\pi} \, e^{i \alpha \hat{\Omega}\cdot \hat{d}}\, \tilde{e}_{ij}^A (\hat{\Omega}) \tilde{e}_{k\ell}^A (\hat{\Omega})\;, \nonumber \\
&& \label{eq58} 
\end{eqnarray}
the overlap reduction functions, Eqs. (\ref{eq13}), (\ref{eq28}), and (\ref{eq29}), can be written as
\begin{equation}
\gamma^M (f) = D^{ij} D^{k\ell}  \Gamma _{ijk\ell}^M (\alpha, \hat{d}\,) \;,
\label{eq45}
\end{equation}
where the superscript $M$ distinguishes polarization modes, $M=T$ (tensor), $V$ (vector), and $S$ (scalar). However, since the following calculations are parallel for all polarization modes, we omit $M=T,V,$ and $S$. In Eqs.\,(\ref{eq44}), (\ref{eq57}), and (\ref{eq58}), the summation about the polarizations is $A=+,\times$ for the tensor mode, $A=x,y$ for the vector mode, and $A=b,\ell$ for the scalar mode, respectively. 

The tensor $\Gamma _{ijk\ell}$ has the symmetric properties,
\begin{equation}
\Gamma _{ijk\ell} = \Gamma _{jik\ell}\;,\;\;\;\;\;\; \Gamma _{ijk\ell} = \Gamma _{ij\ell k}\;,\;\;\;\;\;\; \Gamma _{ijk\ell} = \Gamma _{k\ell ij}\;. \nonumber
\end{equation}
Consequently, thanks to the symmetries, $\Gamma _{ijk\ell}$ can be expanded in tensorial bases as
\begin{eqnarray}
\Gamma_{ijk\ell} (\alpha, \hat{d}\,) &=& C_1 (\alpha) \delta_{ij} \delta_{k\ell} + C_2 (\alpha) (\delta_{ik} \delta_{j\ell}+\delta_{jk} \delta_{i\ell}) \nonumber \\
&+& C_3 (\alpha) (\delta_{ij} \hat{d}_k \hat{d}_{\ell} + \delta_{k\ell} \hat{d}_i \hat{d}_j) \nonumber \\
&+& C_4 (\alpha) (\delta_{ik} \hat{d}_j \hat{d}_{\ell} + \delta_{i\ell} \hat{d}_j \hat{d}_k + \delta_{jk} \hat{d}_i \hat{d}_{\ell} \nonumber \\
&&+ \delta_{j\ell} \hat{d}_i \hat{d}_k) + C_5 (\alpha) \hat{d}_i \hat{d}_j \hat{d}_k \hat{d}_{\ell} \;. \nonumber \\
&& \label{eq46}
\end{eqnarray}

Here we define the contracted quantities of $\Gamma _{ijk\ell}$ with the tensorial bases,
\begin{eqnarray}
q_1 &\equiv& \Gamma_{ijk\ell} \delta^{ij} \delta^{k\ell},\;\;\;\; \nonumber \\
q_2 &\equiv& \Gamma_{ijk\ell} (\delta^{ik} \delta^{j\ell}+\delta^{jk} \delta^{i\ell})\;,\;\;\;\; \nonumber \\
q_3 &\equiv& \Gamma_{ijk\ell} (\delta^{ij} \hat{d}^k \hat{d}^{\ell} + \delta^{k\ell} \hat{d}^i \hat{d}^j)\;,\;\; \nonumber \\
q_4 &\equiv& \Gamma_{ijk\ell} (\delta^{ik} \hat{d}^j \hat{d}^{\ell} + \delta^{i\ell} \hat{d}^j \hat{d}^k + \delta^{jk} \hat{d}^i \hat{d}^{\ell} + \delta^{j\ell} \hat{d}^i \hat{d}^k)\;,\;\;\;\; \nonumber \\
q_5 &\equiv& \Gamma_{ijk\ell} \hat{d}^i \hat{d}^j \hat{d}^k \hat{d}^{\ell} \;. \nonumber \\
&& 
\label{eq47}
\end{eqnarray}
Then, from Eqs.\,(\ref{eq46}) and (\ref{eq47}), $q_1,\cdots,q_5$ can be related to the coefficients $C_1,\cdots,C_5$ by 
\begin{equation}
\left(
\begin{array}{c} 
q_1 \\
q_2 \\
q_3 \\
q_4 \\
q_5 
\end{array}
\right)
=
\left(
\begin{array}{ccccc} 
9 & 6 & 6 & 4 & 1 \\
6 & 24 & 4 & 16 & 2 \\
6 & 4 & 8 & 8 & 2 \\
4 & 16 & 8 & 24 & 4 \\
1 & 2 & 2 & 4& 1
\end{array}
\right)
\left(
\begin{array}{c} 
C_1 \\
C_2 \\
C_3 \\
C_4 \\
C_5 
\end{array}
\right) \;,
\nonumber
\end{equation}
or, inversely,
\begin{equation}
\left(
\begin{array}{c} 
C_1 \\
C_2 \\
C_3 \\
C_4 \\
C_5 
\end{array}
\right)
=\frac{1}{8}
\left(
\begin{array}{ccccc} 
3 & -1 & -3 & 1 & 1 \\
-1 & 1 & 1 & -1 & 1 \\
-3 & 1 & 5 & -1 & -5 \\
1 & -1 & -1 & 2 & -5 \\
1 & 1 & -5 & -5 & 35
\end{array}
\right)
\left(
\begin{array}{c} 
q_1 \\
q_2 \\
q_3 \\
q_4 \\
q_5 
\end{array}
\right) \;.
\label{eq48}
\end{equation}

On the other hand, from Eq.\,(\ref{eq44}), $q_1,\cdots,q_5$ can be explicitly integrated with respect to the propagation direction of GWs over the celestial sphere, by temporarily introducing a coordinate such that
\begin{equation}
\hat{\mathbf{\Omega}} \cdot \hat{\mathbf{d}} = \cos \delta \equiv x, \quad \hat{\mathbf{m}} \cdot \hat{\mathbf{d}} =0, \quad \hat{\mathbf{n}} \cdot \hat{\mathbf{d}} = -\sin \delta, \nonumber
\end{equation}
and by using the integral formulae of spherical Bessel functions $j_n (x)$,
\begin{eqnarray}
\int_{-1}^{1} dx \,e^{i \alpha x}  &=& 2 j_0 (\alpha) \;, \nonumber \\
\int_{-1}^{1} dx \,e^{i \alpha x} (1-x^2) &=& 4 \frac{j_1 (\alpha)}{\alpha} \;, \nonumber \\
\int_{-1}^{1} dx \,e^{i \alpha x} (1-x^2)^2 &=& 16 \frac{j_2 (\alpha)}{\alpha^2} \;, \nonumber \\
\int_{-1}^{1} dx \,e^{i \alpha x} x^2 &=& \frac{2}{3} \biggl[ j_0 (\alpha) -2 j_2 (\alpha)\biggr] \;, \nonumber \\
\int_{-1}^{1} dx \,e^{i \alpha x} x^4 &=& \frac{2}{35} \biggl[ 7 j_0 (\alpha) -20 j_2 (\alpha) +8 j_4 (\alpha) \biggr] \;, \nonumber 
\end{eqnarray}
and the relations between the spherical Bessel functions with different indices,
\begin{eqnarray}
\frac{j_1(\alpha)}{\alpha} &=& \frac{1}{3} \biggl[ j_0(\alpha) + j_2(\alpha) \biggr] \;, \nonumber \\
\frac{j_2(\alpha)}{\alpha^2} &=& \frac{1}{105} \biggl[ 7 j_0(\alpha) +10  j_2(\alpha) + 3 j_4(\alpha) \biggr] \;. \nonumber  
\end{eqnarray}
Then, $C_1,\cdots,C_5$ in Eq.\,(\ref{eq48}) can be expressed in terms of the spherical Bessel functions. The results are
\begin{itemize}
\item{for tensor mode}\\
\begin{eqnarray}
q_1 &=& 0 \;,\;\;\;\;\;\; q_2 = 20 j_0(\alpha) \;,\;\;\;\;\;\; q_3 =0 \;, \nonumber \\
q_4 &=& 40 \frac{j_1(\alpha)}{\alpha} \;,\;\;\;\;\;\; q_5= 20 \frac{j_2(\alpha)}{\alpha^2}\;, \nonumber
\end{eqnarray} 
\begin{equation}
\left(
\begin{array}{c} 
C_1 \\
C_2 \\
C_3 \\
C_4 \\
C_5 
\end{array}
\right)
=\frac{1}{42}
\left(
\begin{array}{ccc} 
-28 & 80 & 3 \\
42 & -60 & 3  \\
0 & -120 & -15 \\
0 & 90 & -15 \\
0 & 0 & 105
\end{array}
\right)
\left(
\begin{array}{c} 
j_0 \\
j_2 \\
j_4  
\end{array}
\right)\;, \nonumber
\end{equation}
\\
\\

\item{for vector mode}
\begin{eqnarray}
q_1&=&0 \;,\;\;\;\;\;\; q_2 = 20 j_0(\alpha) \;,\;\;\;\;\;\; q_3 =0 \;, \nonumber \\
q_4 &=& \frac{20}{3} \biggl[ 2 j_0 (\alpha)-j_2 (\alpha) \biggr] \;,  \nonumber \\
q_5 &=& 20 \biggl[ \frac{1}{15} j_0(\alpha)-\frac{1}{21}j_2(\alpha) -\frac{4}{35}j_4(\alpha) \biggr] \;, \nonumber
\end{eqnarray}
\begin{equation}
\left(
\begin{array}{c} 
C_1 \\
C_2 \\
C_3 \\
C_4 \\
C_5 
\end{array}
\right)
=\frac{1}{42}
\left(
\begin{array}{ccc} 
-28 & -40 & -12 \\
42 & 30 & -12  \\
0 & 60 & 60 \\
0 & -45 & 60 \\
0 & 0 & -420
\end{array}
\right)
\left(
\begin{array}{c} 
j_0 \\
j_2 \\
j_4  
\end{array}
\right)\;, \nonumber
\end{equation}
\\

\item{for scalar mode}
\begin{eqnarray}
q_1&=&30 \biggl( \frac{2+\kappa}{1+2\kappa} \biggr) j_0(\alpha) \;, \nonumber \\
q_2 &=& 60 \biggl( \frac{1+\kappa}{1+2\kappa} \biggr) j_0(\alpha) \;, \nonumber \\  
q_3 &=& \frac{20}{1+2\kappa} \biggl[ (2+\kappa) j_0 (\alpha) +2(1-\kappa) j_2 (\alpha) \biggr] \;, \nonumber \\
q_4 &=& \frac{40}{1+2\kappa} \biggl[ (1+\kappa) j_0 (\alpha) +(1-2\kappa) j_2 (\alpha) \biggr] \;, \nonumber \\ 
q_5&=& \frac{2}{7} \biggl[ 7(4+3\kappa) j_0 (\alpha) +20(2-3\kappa) j_2 (\alpha) \nonumber \\
&& +12(1+\kappa) j_4 (\alpha) \biggr] \;, \nonumber
\end{eqnarray}
\\
\begin{widetext}
\begin{equation}
\left(
\begin{array}{c} 
C_1 \\
C_2 \\
C_3 \\
C_4 \\
C_5 
\end{array}
\right)
=\frac{1}{7(1+2\kappa)}  
\left(
\begin{array}{ccc} 
14(3+\kappa) & -20(3-\kappa) & 3(1+2\kappa) \\
7(1+2\kappa) & 10(1+2\kappa) & 3(1+2\kappa)  \\
0 & 30(3-\kappa) & -15 (1+2\kappa) \\
0 & -15(1+2\kappa) & -15(1+2\kappa) \\
0 & 0 & 105(1+2\kappa)
\end{array}
\right)
\left(
\begin{array}{c} 
j_0 \\
j_2 \\
j_4  
\end{array}
\right)\;. \nonumber
\end{equation}
\end{widetext}
\end{itemize} 

From Eqs.\,(\ref{eq45}) and (\ref{eq46}) together with the traceless property of $D_{ij}$, the overlap reduction function can be written as
\begin{eqnarray}
\gamma (f) &=& \rho_1 (\alpha) D^{ij} D_{ij} + \rho_2 (\alpha) D_{\, \,k}^{i} D^{kj} \hat{d}_i \hat{d}_j \nonumber \\
&&+\rho_3 (\alpha) D^{ij} D^{k\ell} \hat{d}_i \hat{d}_j \hat{d}_k \hat{d}_{\ell}
\;, \nonumber
\end{eqnarray}
with the redefinitions of the coefficients, $\rho_1 (\alpha) = 2 C_2 (\alpha)$, $\rho_2 (\alpha) = 4 C_4 (\alpha)$, and $\rho_3 (\alpha) = C_5 (\alpha)$. The new coefficients are given by 
\begin{itemize}
\item{for tensor mode}\\
\begin{equation}
\left(
\begin{array}{c} 
\rho_1^T \\
\rho_2^T \\
\rho_3^T 
\end{array}
\right)
=\frac{1}{14}
\left(
\begin{array}{ccc} 
28 & -40 & 2  \\
0 & 120 & -20 \\
0 & 0 & 35
\end{array}
\right)
\left(
\begin{array}{c} 
j_0 \\
j_2 \\
j_4  
\end{array}
\right)\;, \nonumber
\end{equation}
\item{for vector mode}\\
\begin{equation}
\left(
\begin{array}{c} 
\rho_1^{V} \\
\rho_2^{V} \\
\rho_3^{V} 
\end{array}
\right)
=\frac{2}{7}
\left(
\begin{array}{ccc} 
7 & 5 & -2  \\
0 & -15 & 20 \\
0 & 0 & -35
\end{array}
\right)
\left(
\begin{array}{c} 
j_0 \\
j_2 \\
j_4  
\end{array}
\right)\;, \nonumber
\end{equation}
\item{for scalar mode}\\
\begin{equation}
\left(
\begin{array}{c} 
\rho_1^{S} \\
\rho_2^{S} \\
\rho_3^{S} 
\end{array}
\right)
=\frac{1}{7}
\left(
\begin{array}{ccc} 
14 & 20 & 6 \\
0 & -60 & -60 \\
0 & 0 & 105
\end{array}
\right)
\left(
\begin{array}{c} 
j_0 \\
j_2 \\
j_4  
\end{array}
\right)\;. \nonumber
\end{equation}
\end{itemize}
Note that the parameter $\kappa$ vanishes in the overlap reduction function, because $F_b$ and $F_{\ell}$ have the same response and $\gamma$ is normalized.


\bibliography{GWB-nonTensor}

\end{document}